\documentclass[preprint,showpacs,preprintnumbers]{revtex4}

\usepackage{graphicx}% Include figure files %
\usepackage{dcolumn}% Align table columns on decimal point %
\usepackage{bm}% bold math %
\usepackage{epsfig}%
\usepackage{epstopdf}%
\usepackage{grffile}%
\usepackage{color}%
\usepackage{colordvi}%
\usepackage{array}
\usepackage{amssymb}%
\usepackage{rotating}%
\usepackage{lscape}%
\usepackage{float}%
\usepackage{footnote}%
\usepackage{wasysym}
\usepackage{gensymb}
 \makesavenoteenv{environmentname}%

\usepackage{hyperref}
\usepackage{cleveref}

\begin{document}

%\linenumbers

\title{Singlet scalar dark matter in the non-commutative space-time:

 a viable hypothesis to explain the gamma-ray excess in the galactic center}

\author{Zahra Rezaei}
\email{zahra.rezaei@yazd.ac.ir}
\affiliation{Faculty of Physics, Yazd University, P.O. Box 89195-741, Yazd, Iran}

\author{S. Peyman Zakeri}
\email{peyman.zakeri@ipm.ir}
\affiliation{School of Particles and Accelerators, Institute for Research in Fundamental Sciences (IPM), P.O.Box 19395-5531, Tehran, Iran}

%\author{Tayebeh Alizadeh}
%\email{}
%\affiliation{	Faculty of Physics, %Yazd University, P.O. Box 89195-741, %Yazd, Iran}

\date{\today}

\begin{abstract}
We explore the non-commutative space-time to revive the idea that gamma-ray excess in the galactic center can be the result of particle dark matter annihilation. In the non-commutative theory, the photon spectrum is produced by direct  emission during this annihilation where a photon can be embed in the final state together with other direct products in new vertices. In the various configurations of dark matter phenomenology, we adopt the most common model known as singlet scalar. Calculating the relevant aspects of the model, we can obtain the photon flux in the galactic center. Comparing our numerical achievements with experimental data reveals that non-commutative space-time can be a reliable framework to explain the gamma-ray excess.% and even (in the future) other indirect signals of dark matter detection.
    
\end{abstract}
\keywords{Non-commutative Space-Time, Dark Matter, Gamma Ray Excess}
\pacs{13.90.+i}
\maketitle

%************************************************************************************************
\section{Introduction}\label{sec:int} 
Current studies on dark matter (DM) in ultra galaxy scales have opened a new window for high energy physics researches. Identifying this unseen non-baryonic matter in the energy density of the universe has provided fundamental physical insights into particle physics, gravity, and even neutrino physics. As there is no candidate in the standard model (SM) of particle physics to constrain the observed DM, various frameworks in the beyond have applied to introduce viable candidate(s) on the model-building. Weakly interacting massive particles (WIMPs) \cite{Gondolo:1990dk, Srednicki:1988ce, Chiu:1966kg, YaserAyazi:2018pea} are found to be the most reliable hypothesis in which DM particles are produced by thermal mechanism called freeze-out. In the other well-motivated scenario, feebly interacting massive particles (FIMPs)\cite{McDonald:2001vt, Hall:2009bx, Ayazi:2015jij, Zakeri:2018hhe} are created non-thermally in a sense through an opposite process to the former case, known as the freeze-in. 

Large efforts have been made to detect DM particles through direct and indirect processes. Direct detection experiments, such as XENON \cite{Aprile:2018dbl}, LUX \cite{Akerib:2016vxi}, SuperCDMS \cite{Agnese:2015nto}, PANDAX-II \cite{Cui:2017nnn} and ect. attempt to detect the nuclear recoil in the scattering of DM particles off target nuclei. On the other hand, DM particles can undergo self annihilation and the resulting products are strongly pursued for purposes of indirect detection. These products could be the SM particles such as electrons, positrons, protons, antiprotons, photons and neutrinos. 

High energy photons are highly considered as the DM signal. Their excess is followed by astronomical instruments and is well measured in the Galactic Center (GC) by the Fermi Large Area Telescope (Fermi-LAT) \cite{Atwood:2009ez}. GC gamma-ray excess may also arise from millisecond pulsars \cite{Bartels:2015aea} and cosmic-rays point sources \cite{Lee:2015fea} but the predominant paradigm which persuasively explains this excess is the annihilation of DM \cite{Bergstrom:1997fj, Ipek:2014gua, Daylan:2014rsa} .

DM annihilation as explanation of gamma-ray excess has been dedicated in a lot of works recently \cite{Abazajian:2010sq, Abazajian:2012pn, Babu:2014pxa, Biswas:2015sva}. In principle, DM candidates may annihilate to the SM particles (usually when they are gravitationally trapped inside high dense regions such as GC or etc.) and then, a hard photon can be described as final-state radiation. Such high energy photons can also arise from the decay of a metastable DM candidate and, like the former scenario, interpreted as secondary particles \cite{Modak:2014vva}. It should be noted that study of DM through its annihilation products like photon can reveal also the information about X-ray signal \cite{Krall:2014dba, Kong:2014gea, Frandsen:2014lfa, Baek:2014qwa}  (observed from the Andromeda galaxy (M31) and Perseus cluster) which is not favored in this paper.  

Different candidates have been suggested as DM particles, such as scalars \cite{Andreas:2008xy, Lu:2016dbc, Kakizaki:2016dza, Guo:2010hq}, fermions \cite{Merle:2013gea, Kim:2008pp, Klasen:2013ypa}, vectors \cite{Hambye:2009fg, Hambye:2008bq}, ect.~\cite{Bhattacharya:2018cgx, Fiaschi:2018rky, Matsumoto:2018acr}. % or some combinations of them \cite{Bhattacharya:2018cgx, Fiaschi:2018rky, Matsumoto:2018acr}. 
The singlet scalar DM model is the simplest one which contains just two free parameters \cite{Guo:2010hq}. In this model, DM particles annihilate to the SM particles via the usual Higgs particle. Since photon does not couple with Higgs (and any other neutral particle) straightly, and the final state consists of photon(s) dose(do) not exist in the tree level, all investigated photon excess have been performed in the higher order of perturbation theory, in the SM. 

Generally, $\gamma-$ or $X-$ ray excesses have been observed in the places containing high dencity of (dark) matter, stronge magnetic fields or both of them. In these places, the usual space-time are not reliable. With this motivation in mind, we consider the non-commutative space-time (NCST) framework to explain the gamma-ray excess in the GC.

%Another important motivation from NCST is that 
In the NCST, photon can be coupled with neutral particles. Therefore, photon couples with Higgs \cite{Batebi:2014lua}, in addition to be the direct product of scalar DM decays \cite{Ettefaghi:2009ai}. %As the aforementioned setup is usually based on the assumption of tree-level emission, the analysis results rely on perturbative theory and has its related uncertainties. 
In this work, we aim to produce the gamma by prompt processes in which DM particles can directly decay to a photon in the NCST. %In this way, the mediated particles do not exit, 
This consideration results in less vertices and also sufficient photon flux in an accurate way. 

%{\color{red}Our calculation and analysis of gamma ray emission is based on $\alpha\beta\gamma$ procedure which will be described in sec \ref{sec:GR}.}   

%Non-commutative field theory (NCFT) is a well-motivated hypothesis in string theory {\color{yellow}(a candidate for explaining gravity)} and {\color{red}has a significance role in particle physics phenomenology \cite{Douglas:2001ba, Martin:2013gma, Abreu:2013vaa}. In the NC space-time, the coordinates are operators and do not obey commutative algebra. Hence, we can construct the gauge theories in this space-time and write the new form of the SM Lagrangian and reach to non-commutative SM (NCSM) \cite{Calmet:2001na, Ohl:2004tn}. No new fields are added but the Feynman rules for the NCSM should be introduced \cite{Melic:2005am, Melic:2005fm}. Subsequently the phenomenological aspects should be considered in the model-building of particle physics. Among the various configurations of DM phenomenology, we adopt the most usual and minimal extension of the NCSM (singlet scalar \cite{Ettefaghi:2009ai}) as our DM model in order to generate the observed gamma-ray excess}.

The article is organized as follows. In section \ref{sec:SDM}, we introduce the construction of the NCST and propose our singlet scalar DM model in this space-time. The GC gamma-ray excess is phenomenologically presented in Sec. \ref{sec:GR}, where we have also described its numerical calculations. Aiming to identify the viable parameter space, we also investigate  the cross section of  DM annihilation and discuss the final implications and results in Sec. \ref{sec:Res}. Concluding remarks and also future points of view are summarized in Sec. \ref{sec:con}.

%*****************************************************************************************************************
\section{Non-Commutative space-time (NCST)}\label{sec:SDM}
In the following, we study the NCST theory and the singlet scalar DM properties in this theory.
\subsection{ THE NCST THEORY}
A significant and fundamental point in the NCST is the commutation relation between the coordinates in the canonical version, i.e. 
\begin{equation}\label{eq01}
[\hat{x^{\mu}} , \hat{x^{\nu}}] = i \theta^{\mu\nu},
\end{equation}
where $\hat{x^{\mu}}$ and $\hat{x^{\nu}}$ read as operators and $\theta^{\mu\nu}$ is a constant, real and anti-symmetric tensor. $\theta^{\mu\nu}$ is the non-commutative (NC) parameter with the length squared dimension ($L^2$) and is related to the NC energy scale  $\Lambda_{NC}$  as $\Lambda_{NC} \approx (\sqrt{|\theta^{\mu\nu}}|)^{-1}$. %Eq.~\ref{eq01} defines a minimal surface,  that means the NCST is contain a prefered direction violating the Lorentz invarience. 
To pass from ordinary space to NC one, commutative fields and the ordinary product between them should be replaced respectively by NC fields and the star product which can be described as
\begin{equation}
(f*g)(x)=\exp(\frac{i}{2}\theta^{\mu\nu}\frac{\partial}{\partial x^{\mu}}\frac{\partial}{\partial x^{\mu}})f(x)g(y)|_{y\rightarrow x},
\end{equation}
where $f(x)$ and $g(y)$ are regular functions on $R^n$. Moving to NCST causes some problems such as charge quantization \cite{Hayakawa:1999yt} and gauge group definition \cite{Hayakawa:1999zf}. There are two approaches to overcome these concerns. The first one is Seiberg-Witten (SW) map whereby the gauge group is the SM gauge group $SU(3)\times SU(2)\times U(1)$ and  NC fields are extended in terms of ordinary ones \cite{Calmet:2001na}. The second approach makes the gauge group larger into $U_*(3)\times U_*(2)\times U_*(1)$, then using Higgs mechanism reaches to the SM gauge group \cite{Chaichian:2001py}. Writing field theory in the NCST (via the aforementioned approaches) causes new features which creates new vertices and corrects the ordinary ones in the SM \cite{Melic:2005fm,Melic:2005am,Aschieri:2002mc,Behr:2002wx}. Antisymmetric tensor $\theta^{\mu\nu}$ has six independent components according to  $\theta^{\mu\nu}=(\theta^{0i},\theta^{ij})$ with $i,j=1,2,3$. Since the unitarity of the theory is violated for $\theta^{0i}\neq 0$ \cite{Gomis:2000zz}, the limit $\theta^{0i}= 0$ is chosen. Also, at the leading order we preserve our calculations up to order $\theta^{\mu\nu}$ (defined as $O(\theta)$  hereafter ). In this paper, we employ SW map and use the calculated vertices in \cite{Batebi:2014lua}.
%############################### 
\subsection{Singlet Scalar DM in the NCST}
\begin{figure}[t]
	\centerline {\includegraphics[width=.70\textwidth]{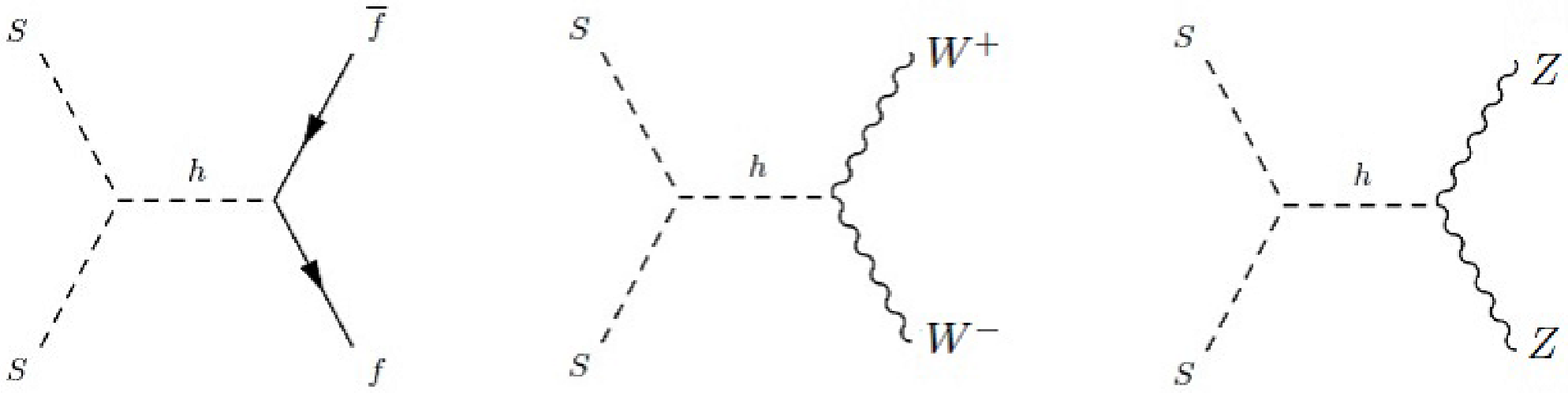} }
	\centerline{ \includegraphics[width=.70\textwidth]{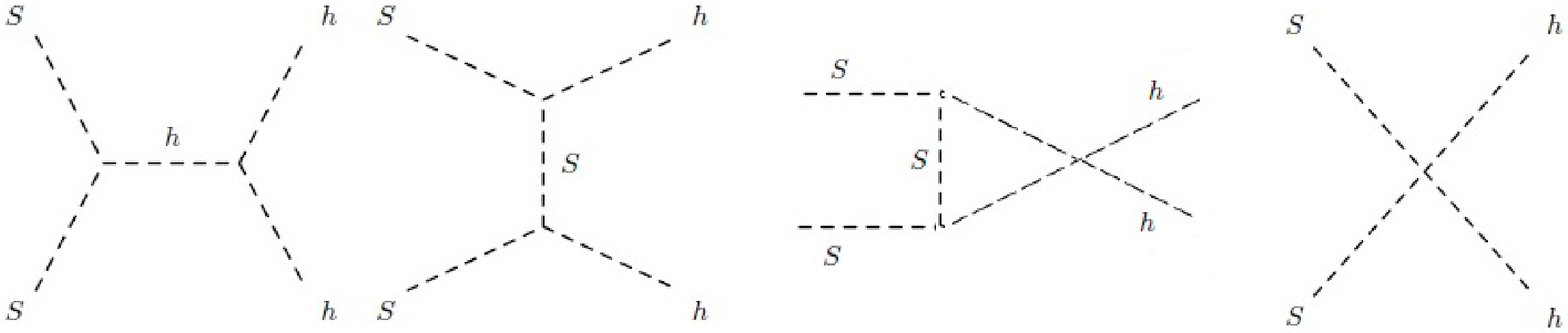}}
	\small\caption{\label{pic1}Tree level diagrams for singlet scalar DM annihilation processes}\end{figure}

One of the most simple and minimal SM extension to describe the particle candidate of DM is made by adding a real singlet scalar particle such that it can reach the equilibrium with the bath particles and plays the role of WIMP DM \cite{Silveira:1985rk,McDonald:1993ex,Burgess:2000yq,Yaguna:2008hd,He:2009yd,Guo:2010hq}. In the sequent model, DM can interact with the thermal soup through the Higgs portal and its stability is usually guaranteed by a discrete $Z_{2}$ symmetry. The
framework of this model is parameterized as:
\begin{equation}
\label{eq1}
\mathcal{L} = \mathcal{L}_{SM} + \frac{1}{2} \partial_{\mu} S \partial^{\mu} S - \frac{m_{0}^{2}}{2} S^{2} - 
\frac{\lambda_{S}}{4} S^{4} - \lambda_{HS} S^{2} H^{\dagger} H.
\end{equation}
In this literature, $S$ denotes the DM and $H$ is the $SU(2)$ Higgs doublet. After spontaneous symmetry breaking, Higgs doublet follows as
\begin{equation}
H = \frac{1}{\sqrt{2}} \left( \begin{array}{c}0  \\v_{H}+h\end{array} \right)\,,
\end{equation}
where $v_H=246$ GeV reads as vacuum expectation value of Higgs, and the Higgs-scalar Lagrangian changes to 
\begin{equation}\mathcal{L} = \mathcal{L}_{SM} + \frac{1}{2} \partial_{\mu} S \partial^{\mu} S - \frac{1}{2} m_S^2 S^{2} - 
\frac{\lambda_{S}}{4} S^{4} - \lambda_{HS} S^{2} h^{\dagger}h-\lambda_{HS}\,\,v_H S^{2}h,\end{equation}
while DM mass is $m_S=\sqrt{m_0^2+\lambda_{HS}v_H^2}$ and $h$ is the SM-like Higgs observed at the LHC. DM annihilation to the SM particles are displayed in Fig.~\ref{pic1} for the singlet scalar DM model. Higgs is the only intermediate particle that connect the singlet scalar DM to the SM particles. Therefore, all final states depend on the  coupling of Higgs and singlet scalar DM that is determined by $\lambda_{HS}$.

As mentioned before, photon could couple to the neutral particles in the NCST and the SM particles of final states could be generated from the mediated photon. The first row processes of Fig.~\ref{pic1} could be possible with mediation of photon in the NCST. The scalar-photon coupling and its results had been investigated for the NCST in Ref.~\cite{Ettefaghi:2009ai}. It has been revealed that cross sections are non-zero just for % for $\theta^{ij}\neq 0$ % limit is zero for mediated photon and  results had been determined for 
  $\theta^{0i}\neq 0$ \cite{Ettefaghi:2009ai}. The limit  $\theta^{0i}=0$ is chosen in the current work, then it is not required to add the mediated photon results. % to our calculations.  %We will explain about it in the following.
   Also, by anomalous three gauge boson couplings that is probable in NCST \cite{Behr:2002wx, Duplancic:2003hg}, the $\gamma\gamma$ final state with mediation of photon is investigated in \cite{Ettefaghi:2012mz}. Since both vertices are of $O(\theta)$ (and therefore the cross section is of $O(\theta^4)$) we put it away, too.
 
In the NCST, the photon emission in the final state of a tree-level channel is possible where Higgs is the mediated particle. In this manner, we have some new vertices such as %that the Higgs boson can couple to %fermion, anti-fermion and photon 
$hhZ$, $hf\bar{f}\gamma$ and $hW^+ W^- \gamma$. Hence, DM can undergo a pair annihilation as $SS\rightarrow f\bar{f}, ZZ, W^+W^-, hh, hZ, f\bar{f} \gamma, W^+W^-\gamma$, where the existed vertices (in the SM) have obtained some corrections from the NCST. %In addition, the new coupling $hhZ$ provides one more pair annihilation as $SS\rightarrow Zh$ that we will consider with aformentioned ones to calculate the cross section of singlet scalar DM annihilation.%   The relevant Feynman diagrams for these processes are depicted in Fig \ref{pic1 , pic2} .

The $SSh$ and $SShh$ couplings could require some corrections in the NCST. To carry out these vertices, one needs to use the Higgs-scalar action after symmetry breaking
\begin{equation}
\mathcal{L}_{HS} =\int d^4 x ( \lambda_{HS} \hat{S}*\hat{S}* \hat{h^{\dagger}}*\hat{h}+\lambda_{HS}\,\,v_H \hat{S}* \hat{S}*\hat{h}),\end{equation}
where the commutative fields ($S$ and $h$) and the ordinary product between them are replaced respectively by the non-commutative fields ($\hat{S}$ and $\hat{h}$) and the star product ($*$). The SW map of the scalar field ($\hat\phi$) can be written as 
\begin{equation}
\hat\phi=\phi+\frac{1}{2}\theta^{\alpha\beta}V_\beta (\partial_\alpha-\frac{i}{2}(V_\alpha\phi-\phi V'_\alpha))+\frac{1}{2}\theta^{\alpha\beta} (\partial_\alpha-\frac{i}{2}(V_\alpha\phi-\phi V'_\alpha))V'_\beta+O(\theta^2),
\end{equation}
where the scalar field ($\phi$) transforms via two different gauge groups with their corresponding gauge fields ($V$ and $V'$). It is straightforward to show that the first order of $SSh$ and $SShh$ couplings in the NCST are of $O(\theta^2)$ and these couplings are removable in the current calculations (It had been shown for $hZZ$ coupling in the NCST \cite{Batebi:2014lua}, too and we put it away).

Extending the NCST with the singlet scalar $S$ embeds three independent parameters as
\begin{equation}
m_S, \lambda_{HS}, \Lambda_{NC},
\end{equation}
whereby we probe the  model parameter space. In the next section, the calculations of gamma-ray excess are presented. 

%*****************************************************************************************************
\section{Gamma-ray excess}\label{sec:GR} 
Our main endeavour in this section is to declare the NCST as a new promising framework to explain the excess of gamma-ray in the energy range 1-3 GeV in the region of GC. A powerful study by Calore, Cholis and Weniger (CCW) \cite{Calore:2014nla} reveals that this GC extended source can be interpreted as DM annihilation. In this regard, latter works have found that such an excess is well fitted with DM interpretation where an annihilation cross section times velocity $<\!\sigma v\!>\propto10^{-27}-10^{-26} \text{cm}^3/s$ \cite{Fermi-LAT:2016uux,TheFermi-LAT:2017vmf,Abramowski:2011hc,Daylan:2014rsa,Coronado-Blazquez:2019puc} is required. In the following, we will investigate whether a WIMP like singlet scalar DM can generate sufficient GC gamma-ray flux in the NCST. The spectrum and amplitude of the gamma-ray signal will be considered for processes in which DM particles annihilate to the SM final states through $s$-channel.

The differential (prompt) photon flux resulting from annihilation of DM particles is given by \cite{Cirelli:2010xx}

\begin{equation}
\frac{d^2{\Phi}}{d{\Omega}\, dE} = \left(\frac{r_{\astrosun}}{8\pi}\right)\left(\frac{\rho_{\astrosun}}{m_{S}}\right)^2 J\cdot \sum_f \langle \sigma v\rangle_{i\to f}\, \frac{dN_{\gamma}^f}{dE}, 
\end{equation}

where $r_{\astrosun}$ is the distance between the Sun and the GC, i.e. $8.33$ kpc and ${\rho}_{\astrosun}$ is the DM density at the solar location with the canonical value $\approx 0.3\, \text{GeV}/\text{cm}^3$ \cite{Bovy:2012tw}. The $J$ factor encodes the effect of matter in line of sight and is given by:

\begin{equation}
J = \int_{\text{l.o.s}} \frac{dr'}{r_{\astrosun}}\left(\frac{\rho(r(r',\theta))}{\rho_{\astrosun}}\right)^2,
\end{equation}

where $r_{\astrosun}$ and $\rho_{\astrosun}$ are added by convention to make $J$ dimensionless. Also, note that $\langle \sigma v\rangle_{i\to f}$ is the cross section of DM ($i$) annihilation to the final states ($f$) and $\, dN_{\gamma}^f/dE\, $ is the energy spectrum of photons produced per one annihilation specifically for the aforementioned final state. The relevant calculations for annihilation cross section are presented in the Appendix. For the integrated flux over a region $\Delta \Omega$, we need to average the $J$ factor over that region:
\begin{equation}
\bar{J}(\Delta\Omega) =  \frac{\int_{\Delta \Omega} J\, d\Omega}{\Delta \Omega},
\end{equation}
 for example if we are interested in $10^{\degree}\times 10^{\degree}$ region around the GC, ${\Delta{\Omega}}= 0.121$ steradians and $\bar{J}=77.7$ \cite{Cirelli:2010xx}, if we assume the Navarro-Frenk-While (NFW) profile for DM \cite{Navarro:1996gj,Klypin:2001xu}.

As mentioned in Ref. \cite{Cirelli:2010xx}, the DM halo profiles are the same for local environment (few pc away from the Earth), but for distances closer to the GC they diverge. Therefore, the calculations for GC region can be sensitive to the choice of the DM profile. According to this gamma emission picking at GC, we assume spherically symmetric and centrally peaked DM halo profile follows NFW as bellow 
\begin{equation}
\rho(r) = \rho_s \frac{(r/r_s)^{-\gamma}}{(1+r/r_s)^{3-\gamma}},
\end{equation}
with the typical scale density $\rho_s= 0.4~\text{GeV}/{\text{cm}^3}$, the radius scale $r_{s} = 20 \text{ kpc}$ and a varied inner slop parameter $\gamma$. In order to calculate the differential flux toward Galactic coordinates, we follow the same choices of parameters as in Ref. \cite{Calore:2014xka}, i.e. our region of interest (ROI) is at Galactic longitude $ |l|\leq 20^{\degree}$ and Galactic latitude $2^{\degree} \leq |b| \leq 20^{\degree}$ where the inner slop parameter is adopted as $\gamma=1.2$ and energy bins (Eq.~(2.2) in Ref \cite{Calore:2014xka}) read as $n_{\text{bins}}=20$ in the range $\left[500 \! \text{MeV}, 500 \! \text{GeV}\right]$.
 
The averaged $J$-factor for this ROI is also given by: $\bar{J}=49.13$. We now calculate the energy spectrum of photons denoted as $\, dN_{\gamma}/dE\, $. Before spectral exploration, a discussion is in order. As it was mentioned earlier in Sec. \ref{sec:int}, we aim to produce prompt photon flux in the leading order of DM annihilation. Thus we consider processes in which photon is produced directly by DM pair annihilation. In this manner, DM particles can annihilate into pure $\bar{b}b\gamma$, $\bar{t}t\gamma$, $\tau^+\tau^-\gamma$ and $W^+W^-\gamma$ channels accompanied with also combinations of them.

To do this, first we start with PPPC4DMID package \cite{Cirelli:2010xx} as a  particle physicist cookbook on calculation of DM signals. Although, PPPC4DMID is a usual instrument to generate photon excess in the SM, the only processes that are considered are DM DM $\rightarrow$ primary primary, where "primary" is a particle of the SM, and photons are produced in the loop levels by the electroweak gauge bosons. Thus, we can not use this package in our presented setup of producing gamma-ray.   
 
In this regard, due to the uncertainty of the GC gamma-ray source, we can perform general log-parabola analysis for the spectrum as \cite{Abazajian:2012pn,Abazajian:2014fta}
\begin{equation}\label{eq:spec}
\frac{dN_{\gamma}}{dE}=N_0(\frac{E}{E_b})^{-(\alpha+\beta\log{(E/E_b)})},
\end{equation} 
where $\alpha$ and $\beta$ are halo parameters and $E_b$ is an arbitrary energy scale. $N_0$ is the number of photons per square centimetre per second per steradian and denotes the possibility of photon emitted per annihilation event. As we produce photon promptly per annihilation channel through its coupling with the Higgs in the NCST, thus, neglecting higher orders of perturbation, we set $N_0=1$ in our formalism.

Now,  the  value of $\alpha$ and $\beta$ parameters ought to be determined. PYTHIA 6.4 \cite{Sjostrand:2006za} is usually used to simulate SM processes that gives rise to the spectrum of photons (as well as other particle spectra). In this case, gamma can be described as final-state radiation of charged particles (e.g. bremsstrahlung spectrum) or neutral ones (e.g. decays of $\pi^0$). In the case of DM interpretation, the aforementioned particles also generate from DM annihilation or decay \cite{Abazajian:2010sq}. As %PYTHIA most emphatically does not simulate the DM profile in the universe and 
our model (singlet scalar DM in the NCST) involves new elementary couplings that generates a different kind of photon flux, we can not utilize PYTHIA and need to calculate the matrix element for photon emission manually. % as there aren't any other packages available at present.
 We will do this in Sec. \ref{sec:Res} and investigate the best fit of our model in terms of the halo parameters and contributing channels.   

Regarding the energy spectrum of photons, it should be noted here that we can employ various spectral models (e.g. one with exponential cutoff \cite{Abazajian:2012pn}) but we have found the logarithmic form (Eq. \ref{eq:spec}) with the best consistency. On the other hand, we should emphasize that $\, dN_{\gamma}/dE\, $ is a property of the annihilation process and does not depend on the halo density profile. Hence, one may also choose Einasto \cite{Graham:2005xx,Oman:2015xda} or $\alpha\beta\gamma$ profiles \cite{Abazajian:2012pn,Abazajian:2014fta}.

In the following, we investigate the model parameter space with DM annihilation cross section and the generation of gamma-ray excess is inquired. 

%**********************************************************************************************************
\section{Results and Disscusion}\label{sec:Res} 
The viability of the procedure introduced in previous sections is tested over a range of model parameters space in this section.

%Our results divide to three part,
%In the following, we turn our attention to the case depending on the coupling between DM and Higgs portal, $\lambda_{HS}$. 
Referring to \cite{Cline:2013gha}, we divide DM mass range into several intervals. The adopted mass range of DM determines the relevant processes through which DM can be annihilated. Our results are best behaved in three intervals. The first interval is around the weak gauge bosons masses ($80-100$ GeV), the second one is selected around Higgs mass ($120-170$ GeV which is smaller than the top quark mass) and the third region is bigger than the top quark mass ($>173$ GeV). In all calculations, the light fermions ($e, \mu, \nu_i, u, d, s$ and $c$) are ignored.  %and depicted in Fig. ? in the logarithmic scale. 
% Varying $\lambda_{HS}$ from 0.032 to 0.13, we start with Fig. a where DM mass changes as 80-100 GeV. The contributing processes and $\Lambda$ and $E_\gamma$ are same as Fig a. Having a small value in feeble coupling, DM cross section features a sharp increment with appropriate values of $m_S$ as 85, 90 and 95 GeV. With similar consideration of Fig b, including relevant processes and the fixed parameters, Fig 2b shows still an increment of cross section in terms of $\lambda_{HS}$ embedding in the interval 0.032-0.26. In the final case, we study DM candidate off large mass (similar to Fig 1c) with appropriate values of the coupling which differs as 0.032 to 1.5. The behavior of $<\sigma v>$ is now has the tightest change in this range and increase with a smaller gradient in large values of DM mass.

We explore the parameter space consistent with the phenomenological predictions of thermal averaged cross section times velocity of DM annihilation, $<\!\sigma v\!>$, and differential photon flux from DM annihilations, $E^2~d\Phi/dE$. To find the correlation between variables $\lambda_{HS}, m_S$ and $\Lambda_{NC}$, the scatter points 
 $(\lambda_{HS}, m_S)$ and $(\Lambda_{NC},m_S)$ are investigated.
%############################
\begin{figure}[!htp]
	\centerline {
\includegraphics[width=.5\textwidth]{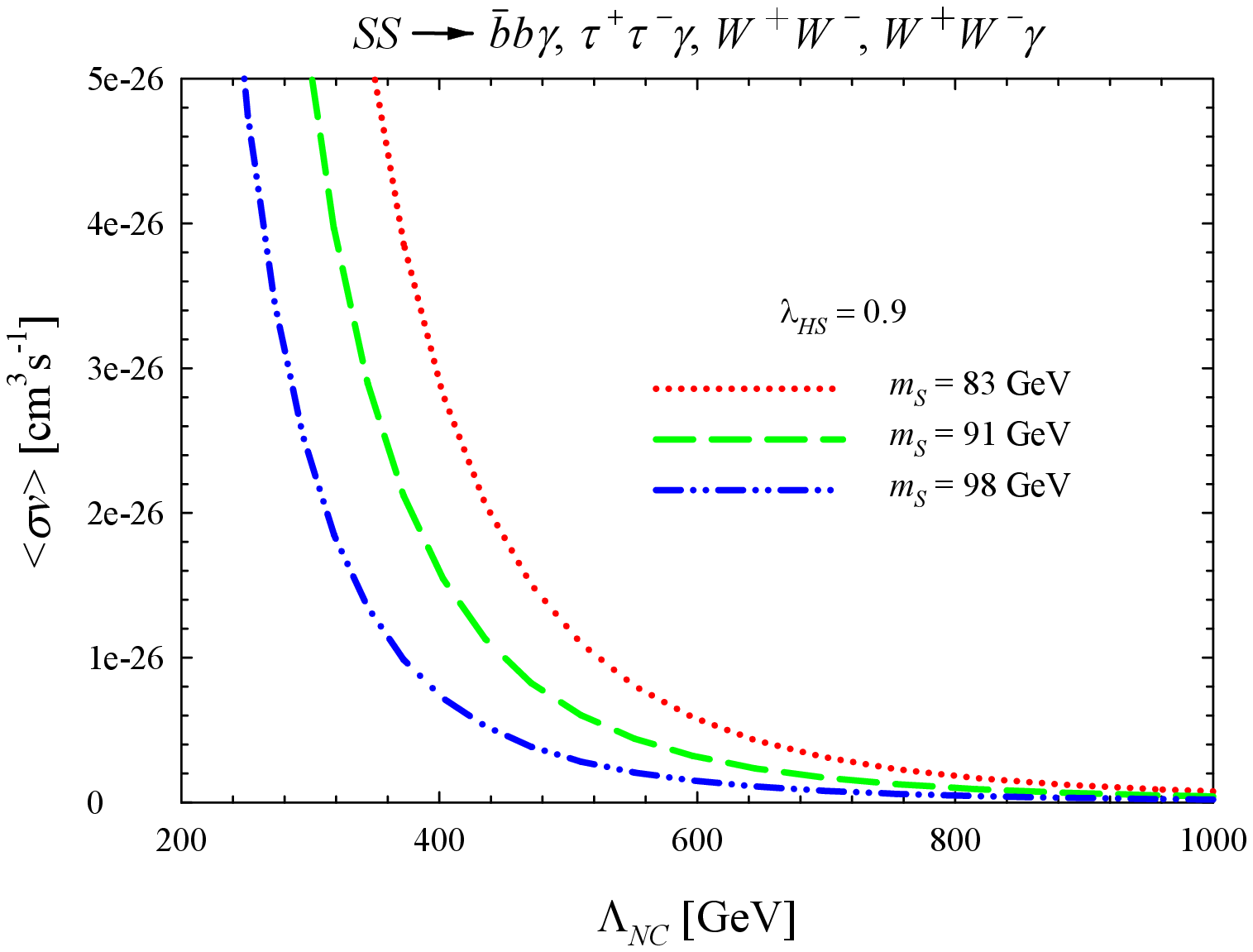}
\includegraphics[width=.5\textwidth]{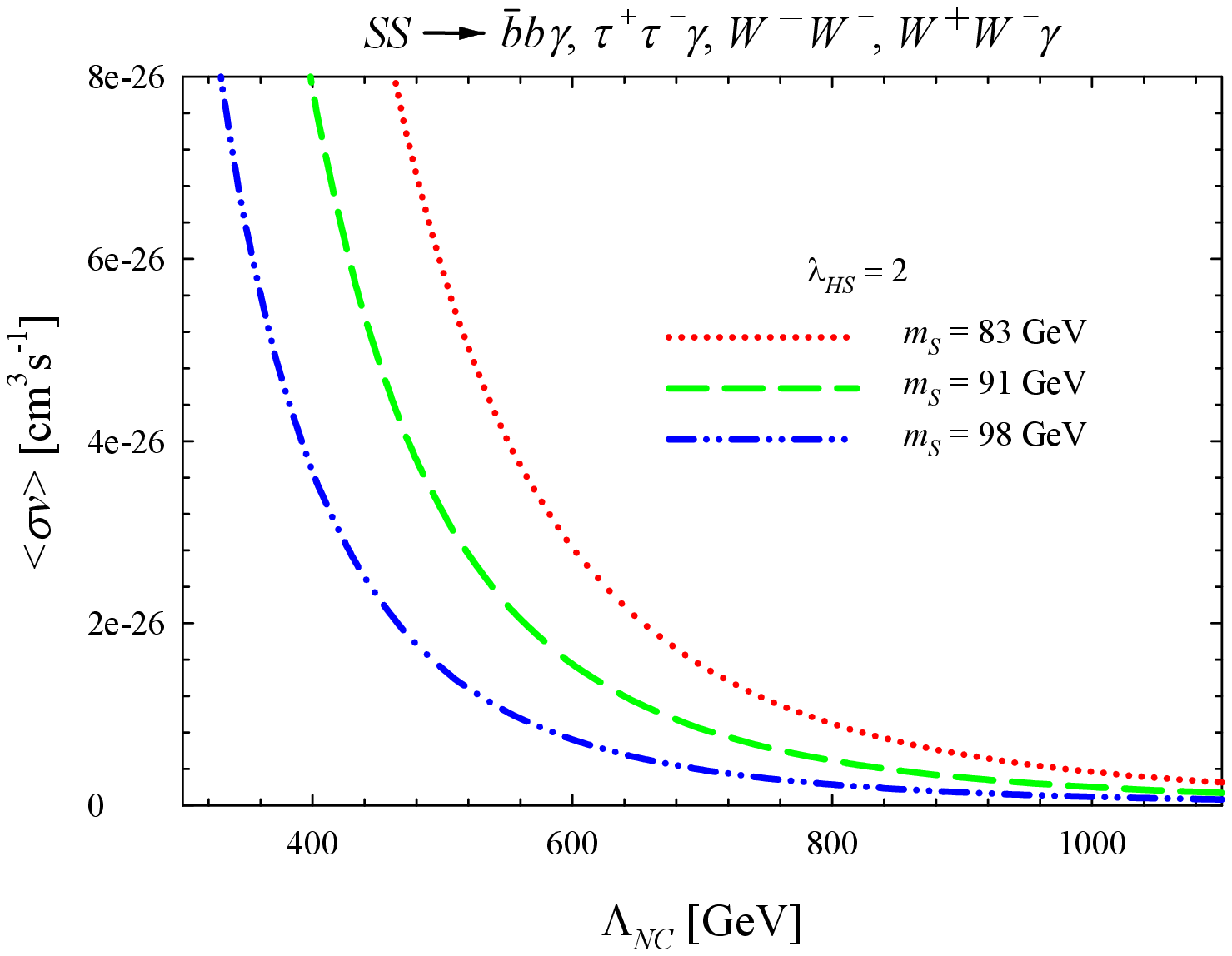}}
\centerline {
 \includegraphics[width=.5\textwidth]{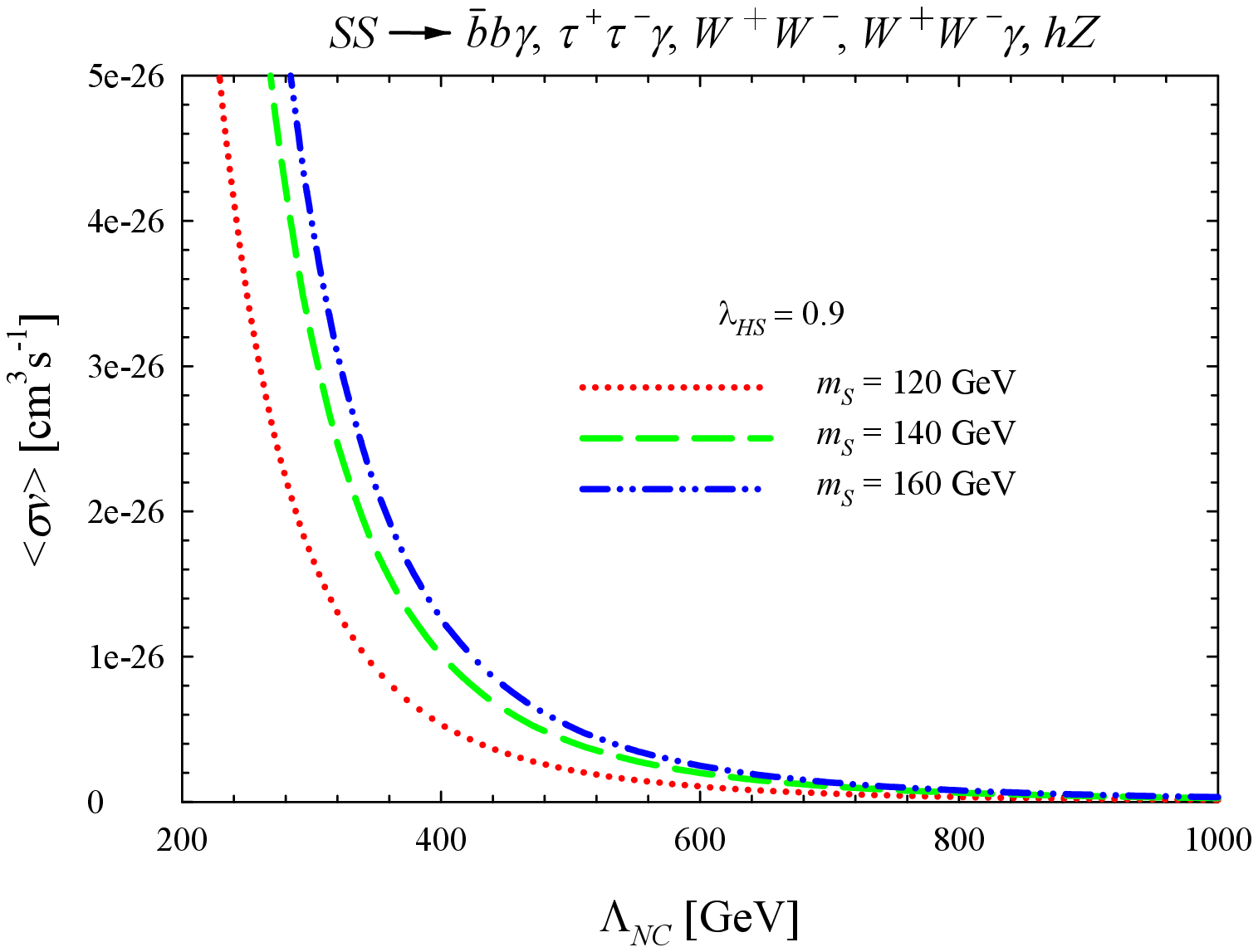}
\includegraphics[width=.5\textwidth]{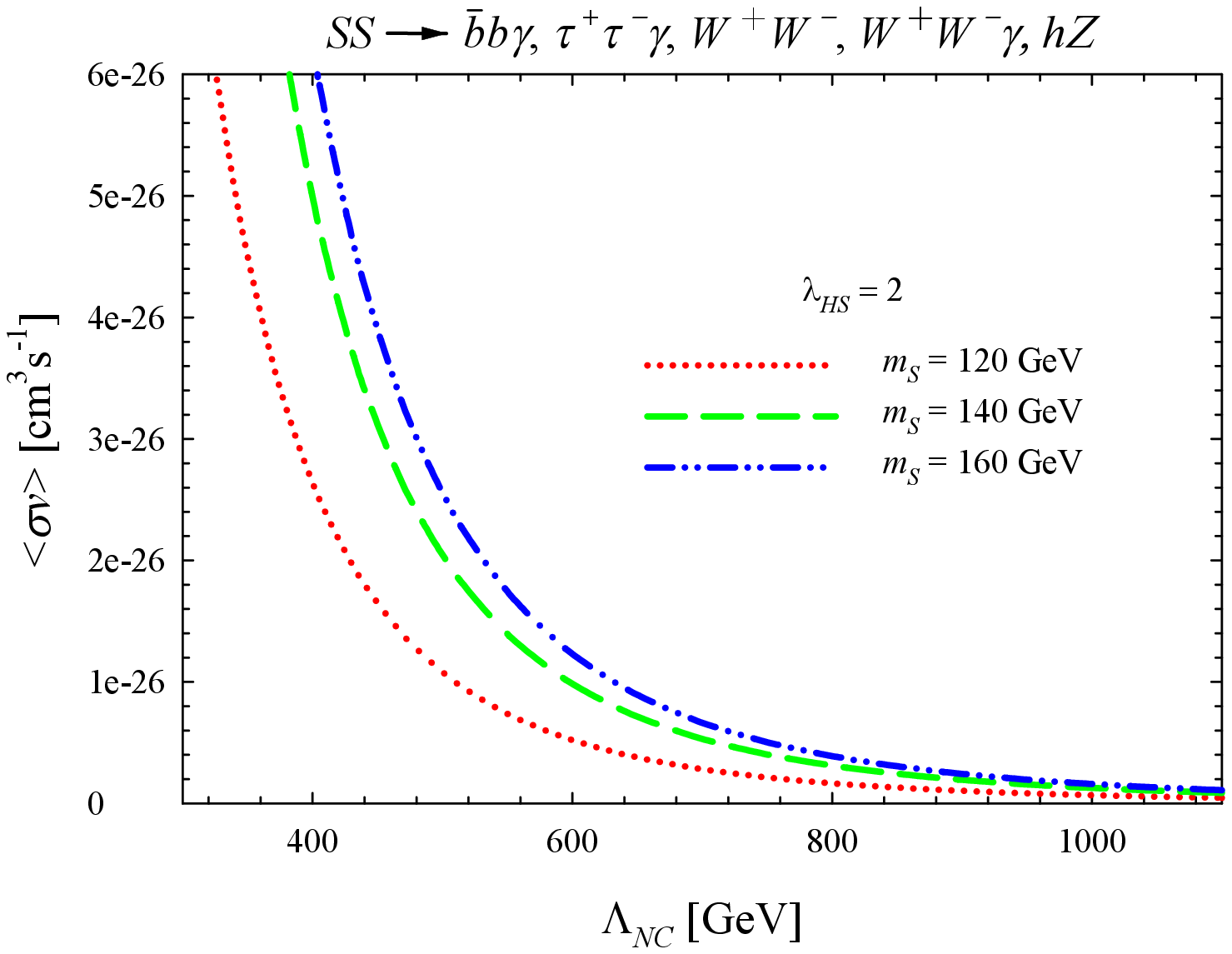}}
\centerline {
\includegraphics[width=.5\textwidth]{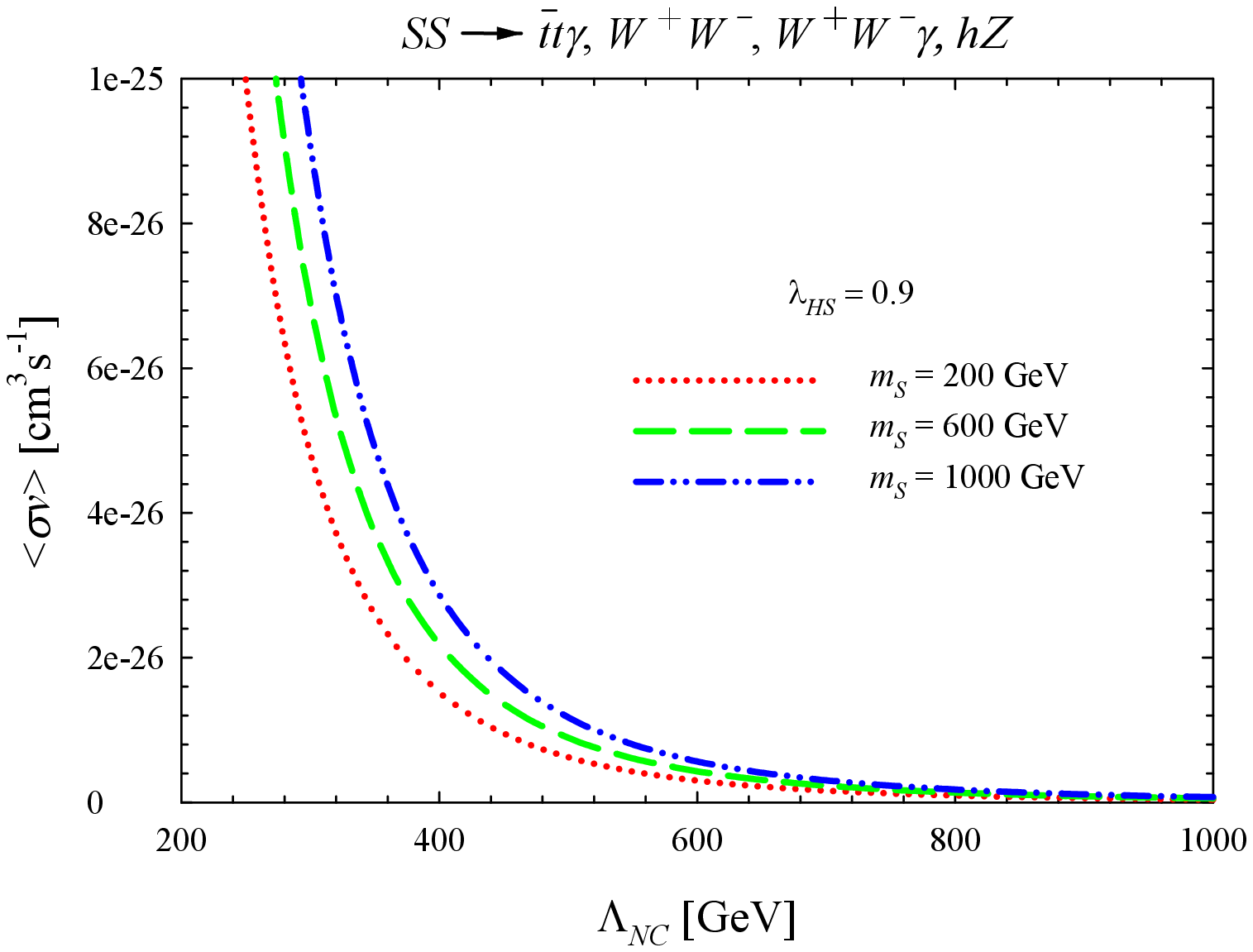}
\includegraphics[width=.5\textwidth]{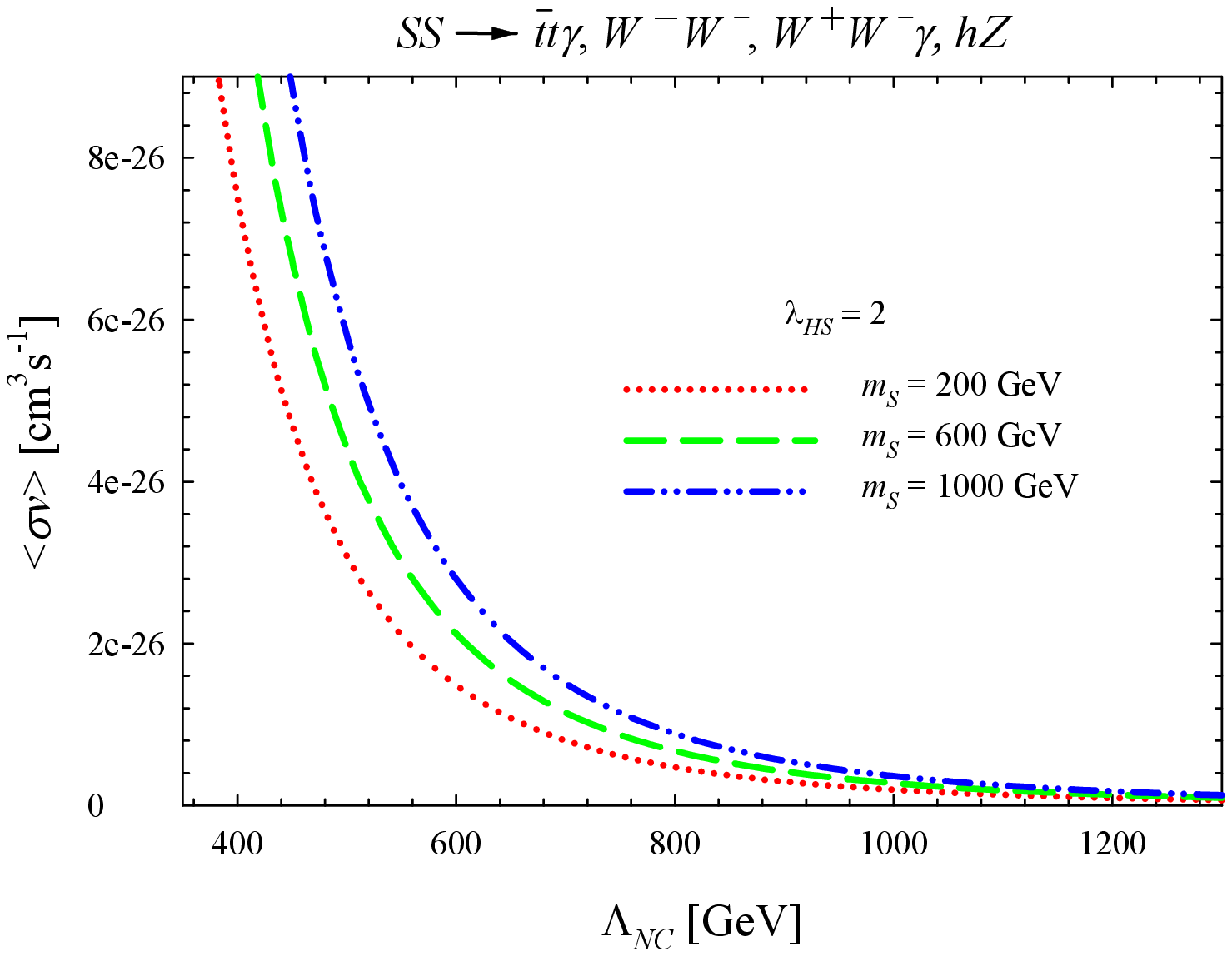}} 
	\small\caption{\label{svL} DM annihilation cross section in terms of NC energy scale. In this figure, all contributing channels are considered.  For different values of $m_S$, we set $\lambda_{HS}=0.9$ (left panel) and $\lambda_{HS}=2$ (right panel).}\label{fig.SVL1}
\end{figure}
%##########################

Figure~\ref{fig.SVL1} describes $<\!\sigma v\!>$ versuse $\Lambda_{NC}$. It shows that $<\!\sigma v\!>$ decreases for all DM mass intervals. %The adopted mass range of DM determines the relevant process through which DM can be annihilated. 
 The final states of DM annihilation are chosen as shown in the above of each plot of Fig.~\ref{fig.SVL1}, according to the relevant DM mass. In the first region, the DM annihilation reads as $SS\rightarrow f\bar{f}, f\bar{f}\gamma, ZZ, W^+W^-, W^+W^-\gamma$ where fermion $f$ indicates $\tau$ lepton and $b$ quark. In pure NCST, $SS\rightarrow f\bar{f}$ is zero and $SS\rightarrow ZZ$ is of order ($>\theta^2$), therefore they don't play any role in our considerations. In the second region, DM particles undergo an annihilation through $SS\rightarrow f\bar{f}, f\bar{f}\gamma, ZZ, W^+W^-, W^+W^-\gamma, Zh, hh$ where only the final states $f\bar{f}\gamma, W^+W^-, W^+W^-\gamma, Zh$ (which $f$ is still to be defined as $\tau$ and $b$ particles) are adopted in NC hypothesis. The threshold energy to produce $Zh$ is around 108 GeV, therefore we have chosen the DM mass range few GeVs bigger and so nearer the Higgs mass. %It seems the new channel ($SS\rightarrow Zh$) For heavy DM masses (> 173 GeV), the $b$ and $\tau$ are nachiz and $top$ quark is the most important fermion to consider in the final state. Therefore, the DM annihilation channels are  $t\bar{t}\gamma, W^+W^-, W^+W^-\gamma, Zh$. 
The third analysis devotes to the heavy massive DM where its mass varies from 173 GeV to 1 TeV. The contributing processes for DM annihilation are the same as former cases except that $f$ indicates top quark here. As shown in Fig.~\ref{fig.SVL1}, increasing DM mass or$/$and Higgs-DM coupling ($\lambda_{HS}$), the desired cross section acquires in the larger $\Lambda_{NC}$. %DM annihilation cross section $<\!\sigma v\!>$ changes with respect to NC scale $\Lambda_{NC}$. %There are two regims, one is decreasing with respect to  $\Lambda{NC}$ and another is increasing. The increment appear when the $SS\rightarrow Zh$ added to the signals. .......... %It shows obvously 
%@@@@@@@@@@@@@@@@@@@@@@@
%Fig ?? b depicts this consideration via proper values of the coupling between DM and Higgs where we see a continuous decrement of annihilation cross section in the whole ranges of DM mass. Three influencing values of $\lambda_{HS}$ are considered and the result is shown in Fig c. It is obvious from this Fig that DM annihilation cross section decreases in a same way of previous intervals as far as it vanishes approximately in $m_S=1$ TeV.

%igs ?? show our results in the log-scale for proper parameter spaces. In Fig ?? a, DM mass ranges from 80 GeV to 100 GeV, so  Regarding different values of $\lambda_{HS}$ and flxed values of relevant parameters $\Lambda=1$ TeV and $E_\gamma=3$ GeV, the behavior of $<\sigma v>$ features a small increment in low masses (until 82.5 GeV) and then decreases as $m_S$ reaches 100 GeV. The other allowed parameter space is made on the basis where DM mass gets values form 100 GeV to 173 GeV. In this regard, 
%@@@@@@@@@@@@@@@@@@@@@@@@@@@@
\begin{figure}[!htb]
	\centerline {\includegraphics[width=.50\textwidth]{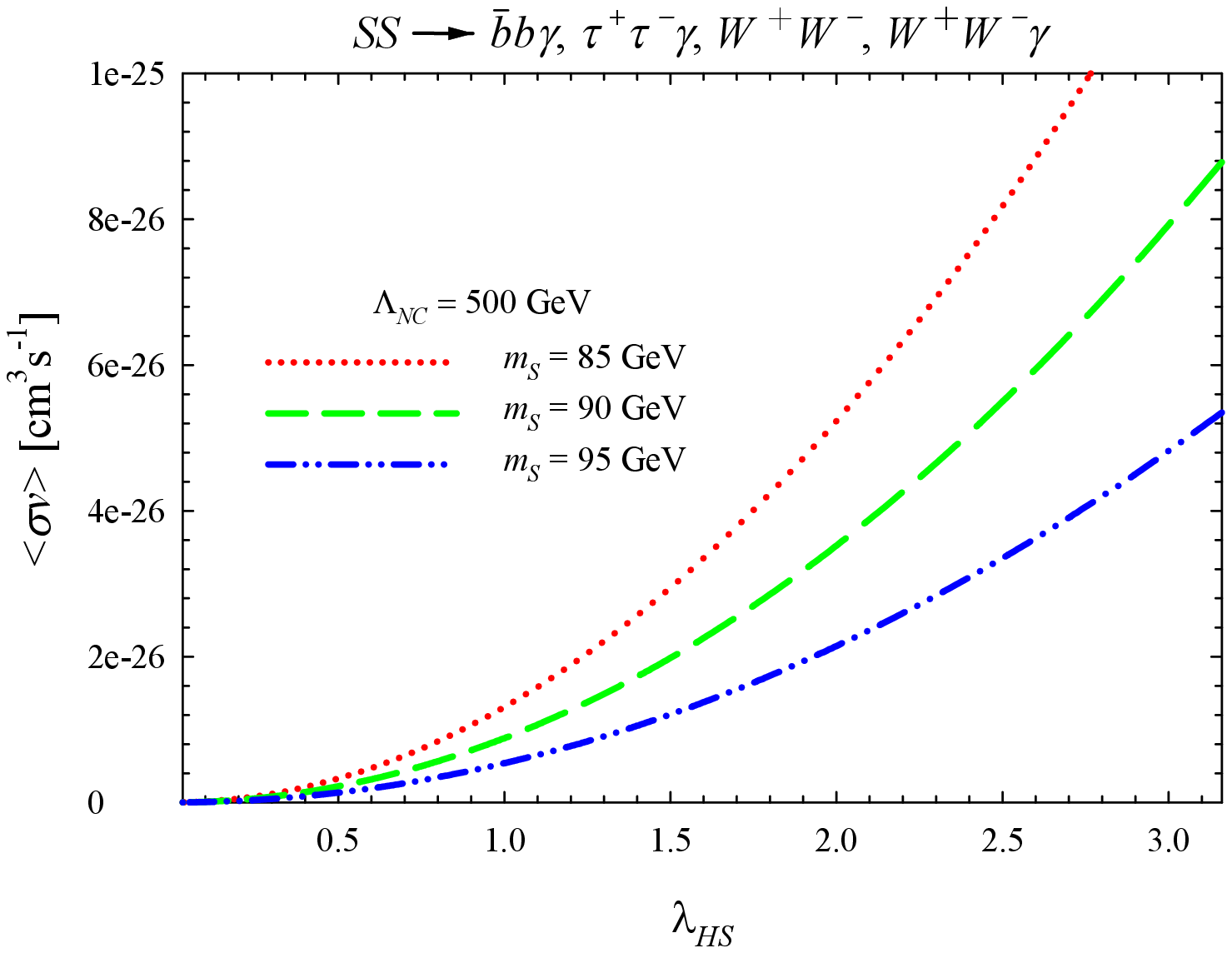}
\includegraphics[width=.50\textwidth]{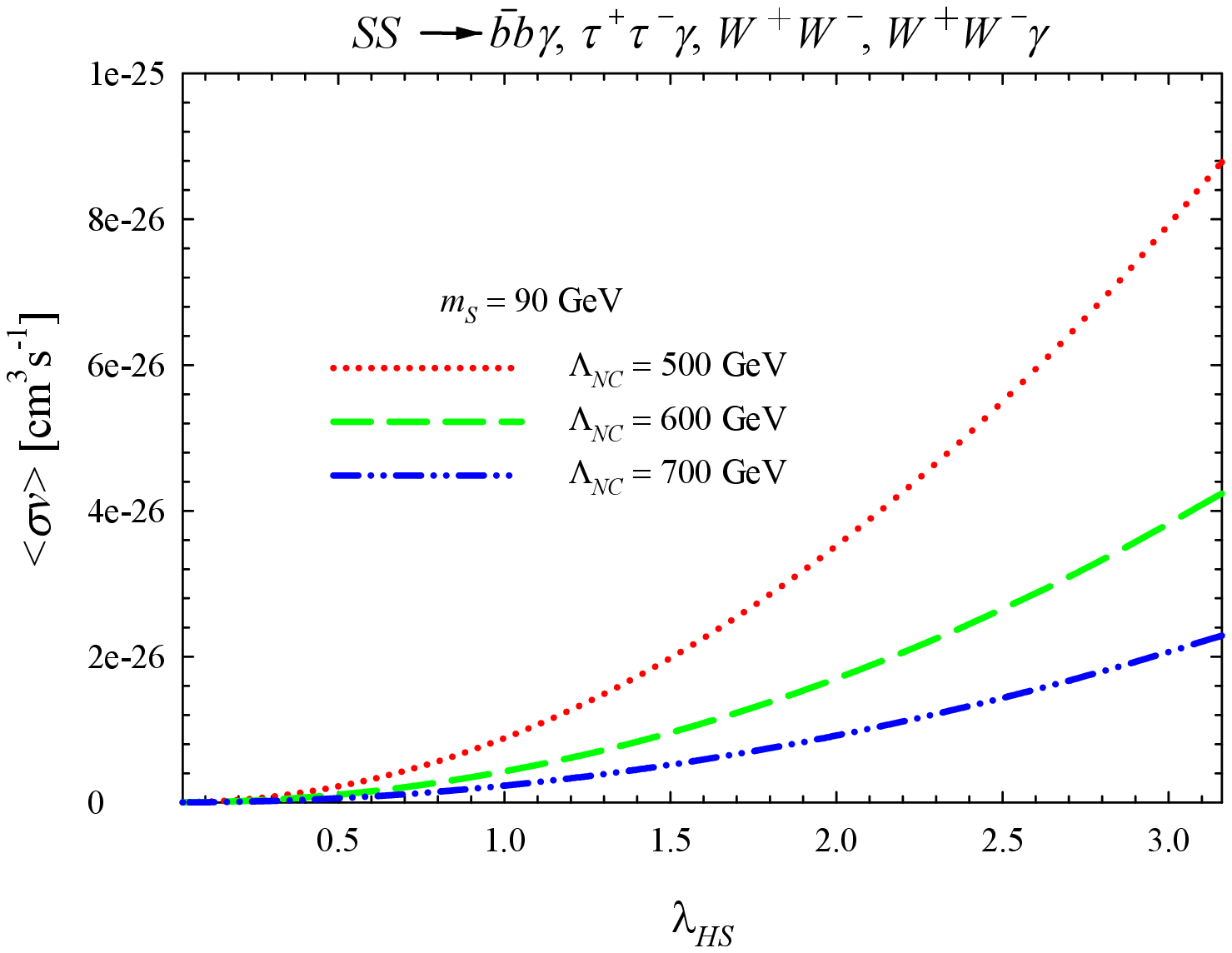}}
 \centerline{\includegraphics[width=.50\textwidth]{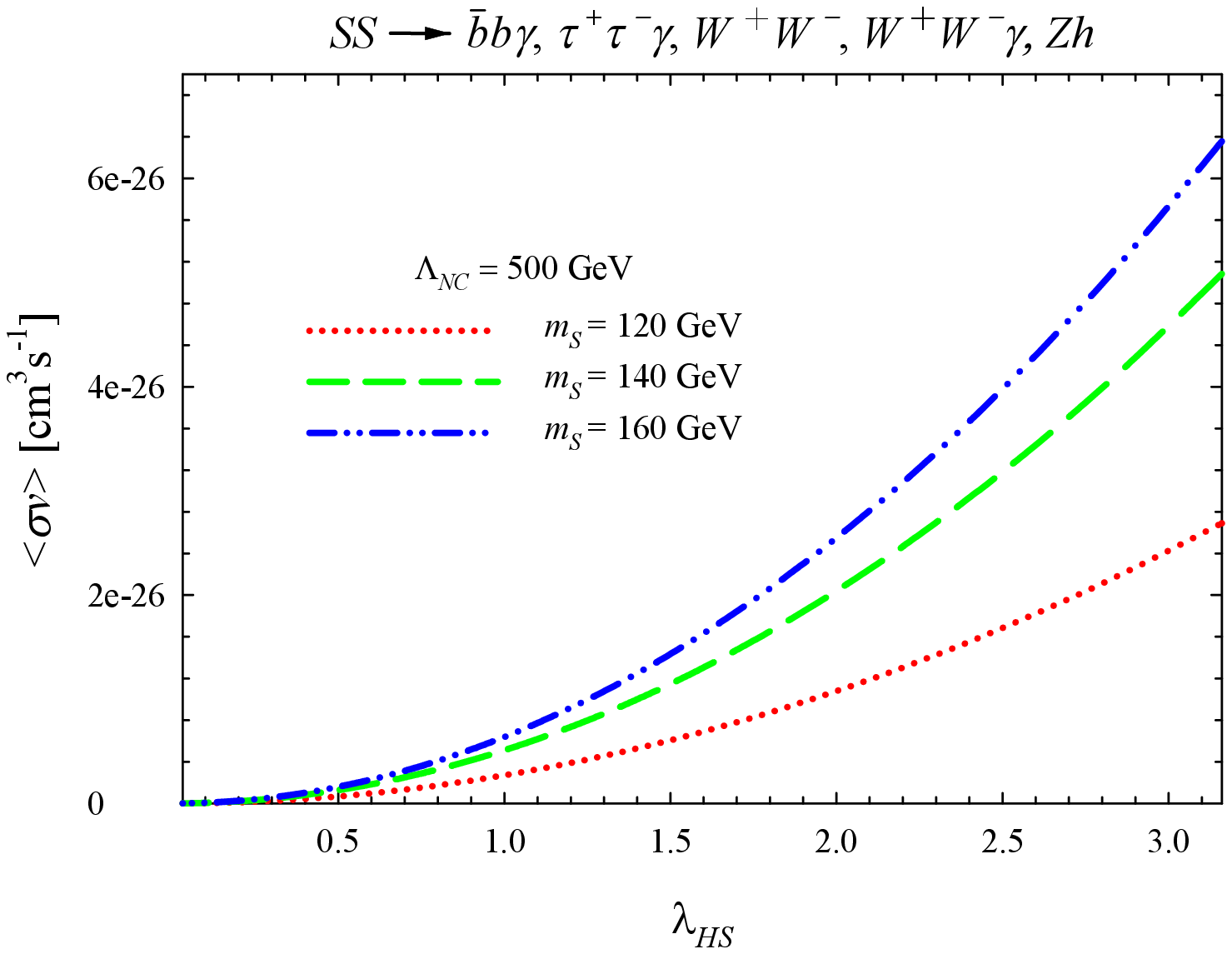}
\includegraphics[width=.50\textwidth]{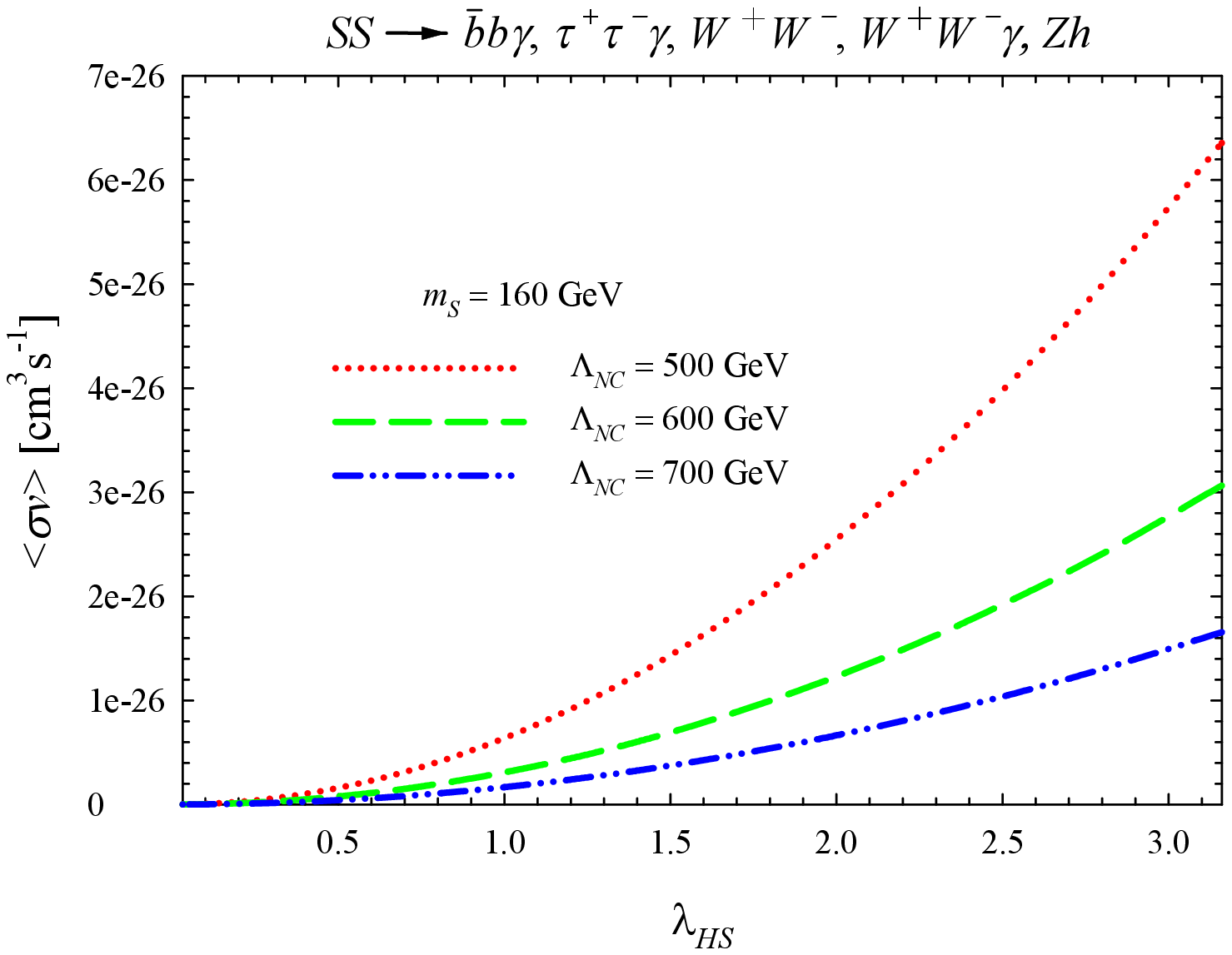}}
\centerline{\includegraphics[width=.50\textwidth]{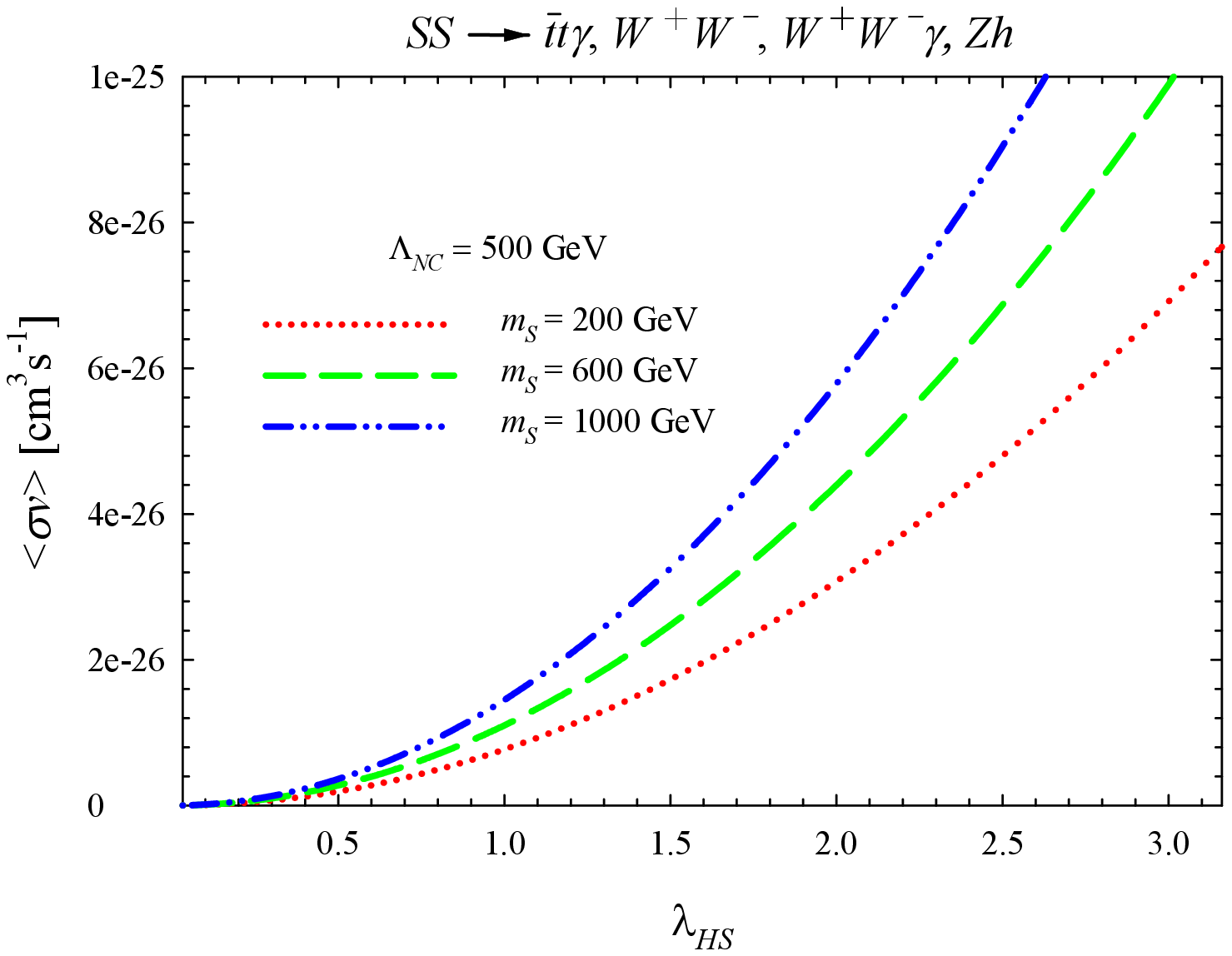}
\includegraphics[width=.50\textwidth]{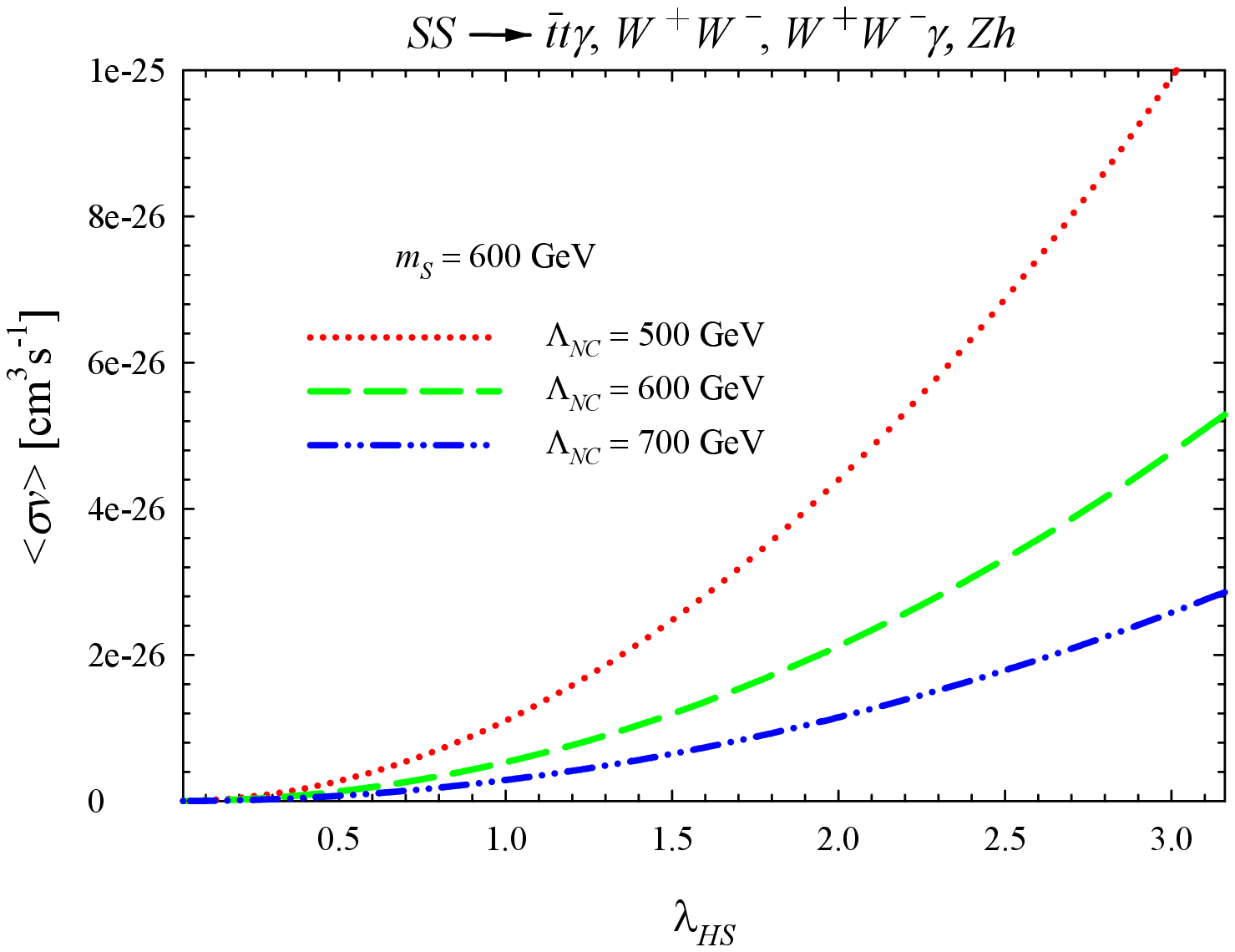}}
	\small\caption{DM annihilation cross section in terms of its coupling. In this figure, all contributing channels are considered. In the left panel, we set $\Lambda_{NC}=500$ GeV for different values of $m_S$ and in the right one corresponding values of $m_S$ are adopted for different choices of $\Lambda_{NC}$.}
\label{fig.SVlHS}
\end{figure}
%&&&&&&&&&&&&&&&&&&&&&&&&&&&&&&&&&
The DM annihilation cross section $<\!\sigma v\!>$ with respect to $\lambda_{HS}$ is depicted in Fig.~\ref{fig.SVlHS}. Three DM mass regions are defined again. Left  plots are depicted for the fixed $\Lambda_{NC}$ and three different masses in each region and the fixed mass with different $\Lambda_{NC}\!$ s are shown in the right side. As it shows, the cross section increases in terms of $\lambda_{HS}$. 
%&&&&&&&&&&&&&&&&&&&&&&&&&&&&&&&&&&
\begin{figure}[!htb]
	\centerline {
\includegraphics[width=.50\textwidth]{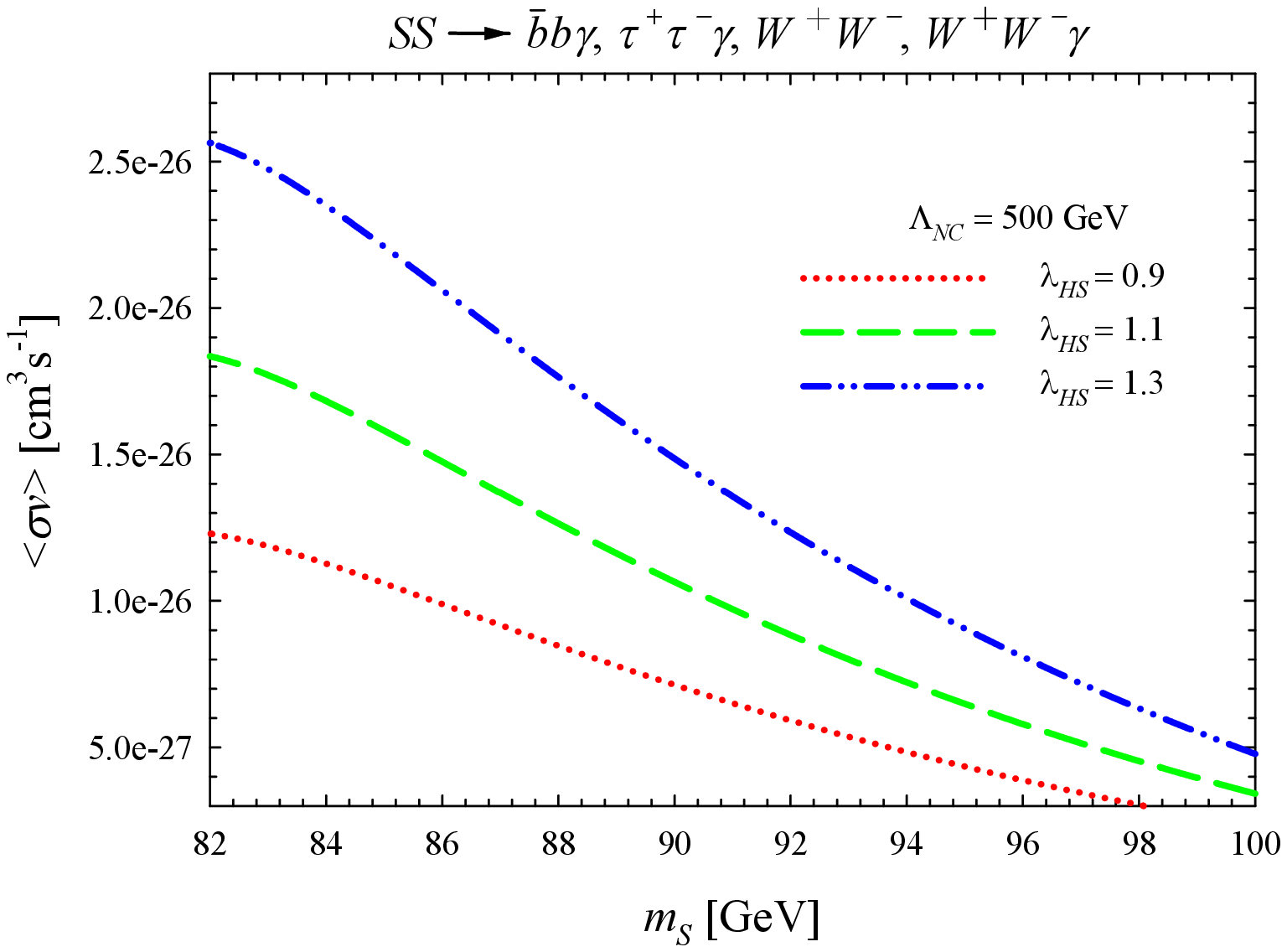}
\includegraphics[width=.50\textwidth]{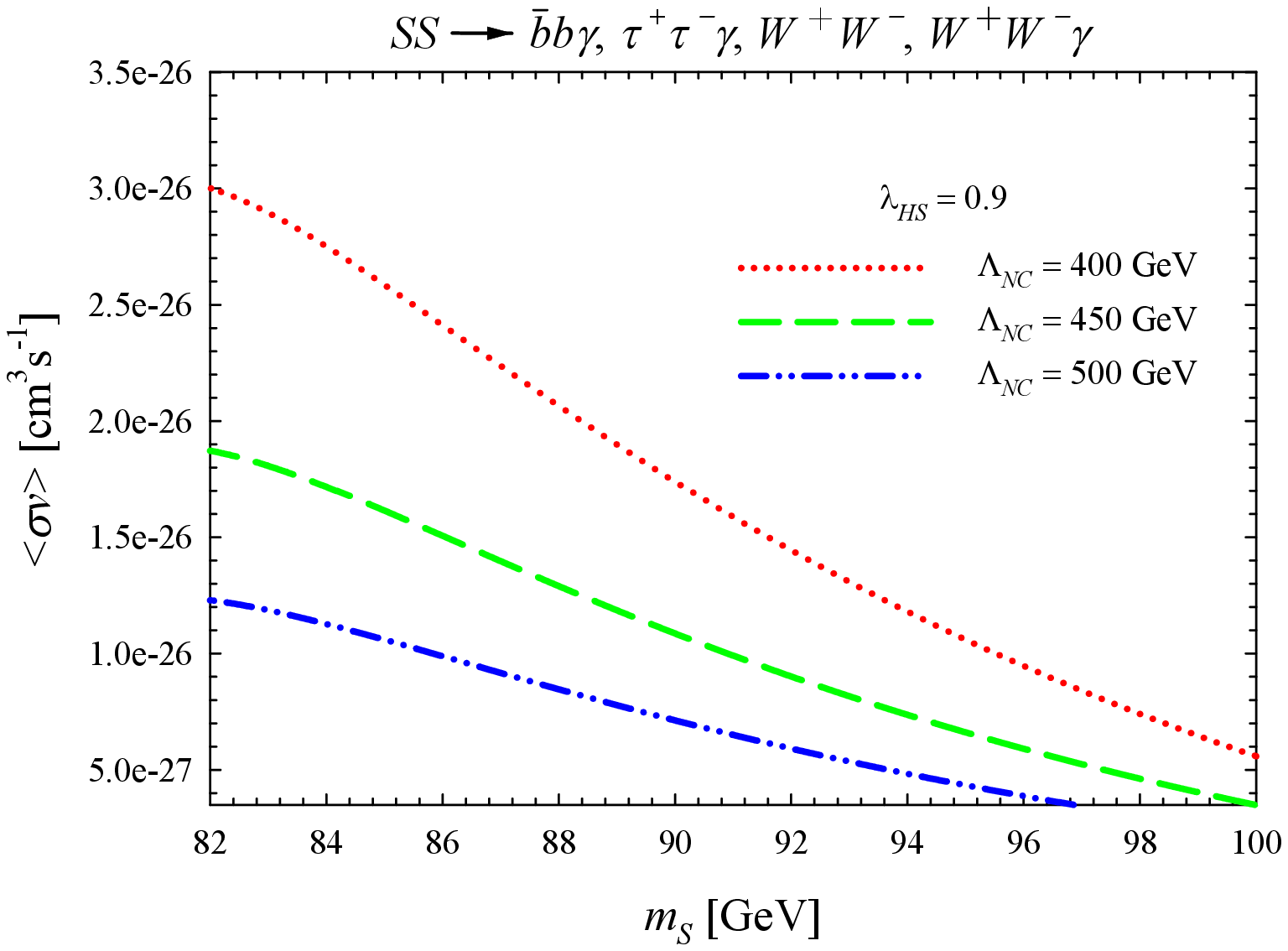}}
\centerline{
 \includegraphics[width=.50\textwidth]{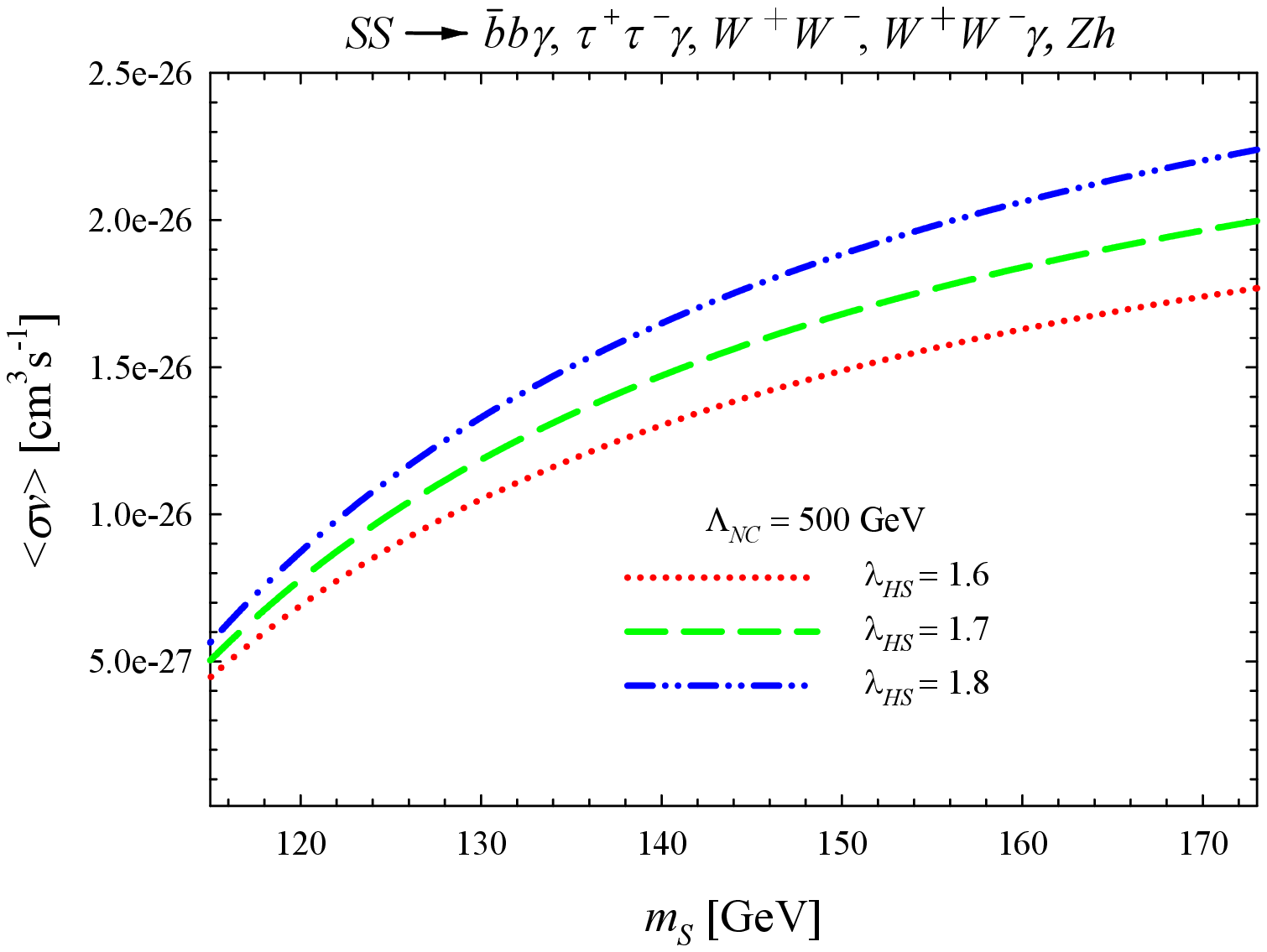}
\includegraphics[width=.50\textwidth]{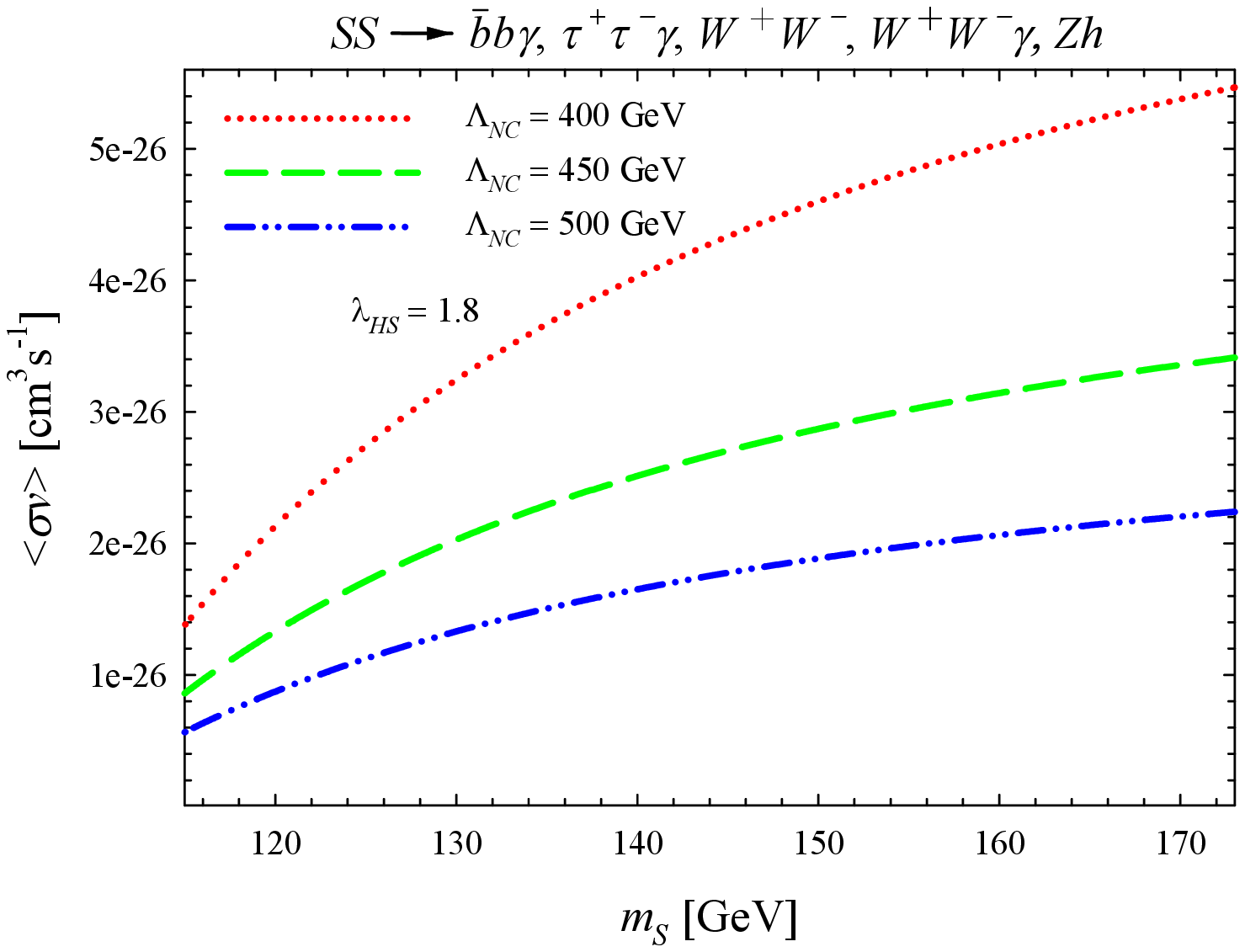}}
\centerline{\includegraphics[width=.50\textwidth]{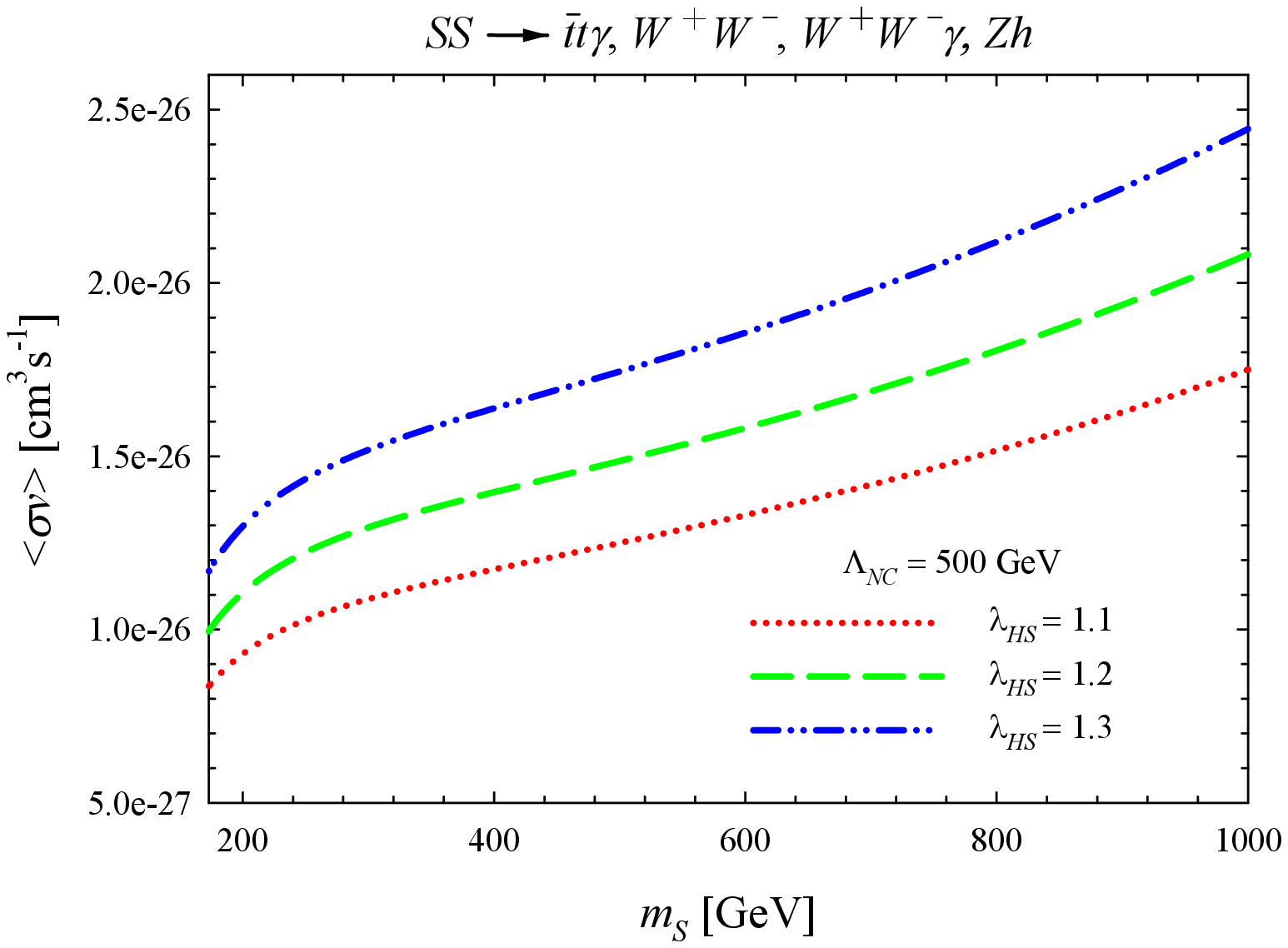}
\includegraphics[width=.50\textwidth]{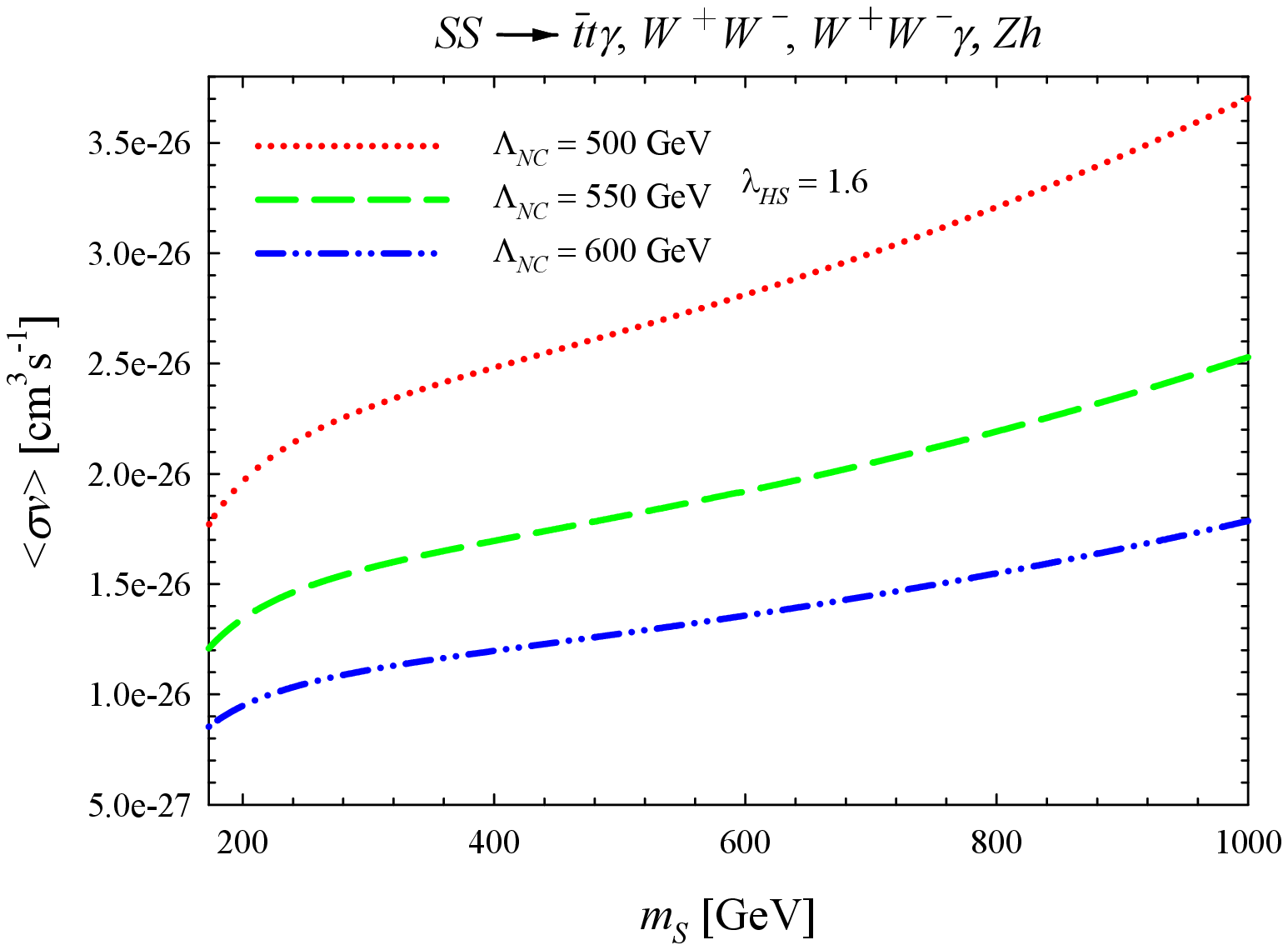} }
	\small\caption{DM annihilation cross section in terms of its mass. In this figure, all contributing channels are considered. In the left panel, we set $\Lambda_{NC}=500$ GeV for different values of $\lambda_{HS}$ and in the right one corresponding values of $\lambda_{HS}$ are adopted for different choices of $\Lambda_{NC}$.}\label{fig.svm1}
\end{figure}

Figure~\ref{fig.svm1} describes $<\!\sigma v\!>$ versus $m_S$ for three regions of DM mass. For lighter DM particles (in the first region), $<\!\sigma v\!>$ decreases with respect to the mass but for the heavier ones, where the new $hhZ$ coupling is appeared from the NCST effects, the slow increment in cross section behaviour is seen. 

\begin{figure}[!htb] 
	\centerline {
\includegraphics[width=.30\textwidth]{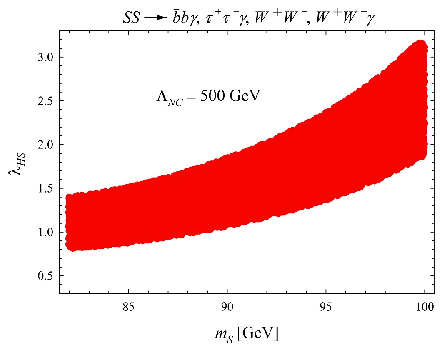} 
 \includegraphics[width=.30\textwidth]{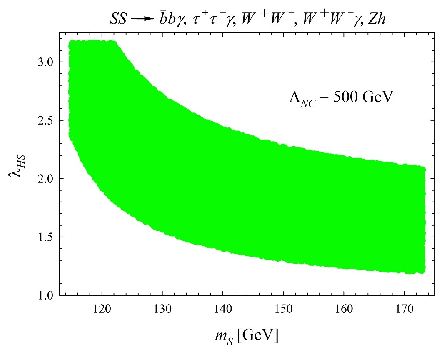}
\includegraphics[width=.30\textwidth]{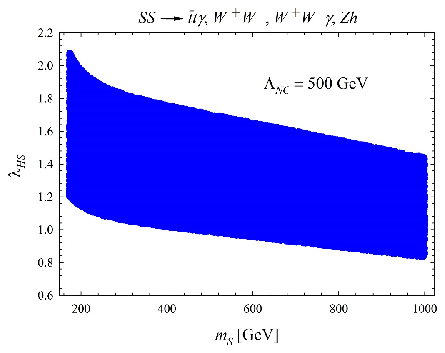}}
	\small\caption{The allowed region in the ($\lambda_{HS},m_S$)-parameter space where $\Lambda_{NC}=500$ GeV. The panels are labeled by their corresponding channels.}\label{fig.scatter-point1}
\end{figure}

\begin{figure}[!htb]
	\centerline {
\includegraphics[width=.30\textwidth]{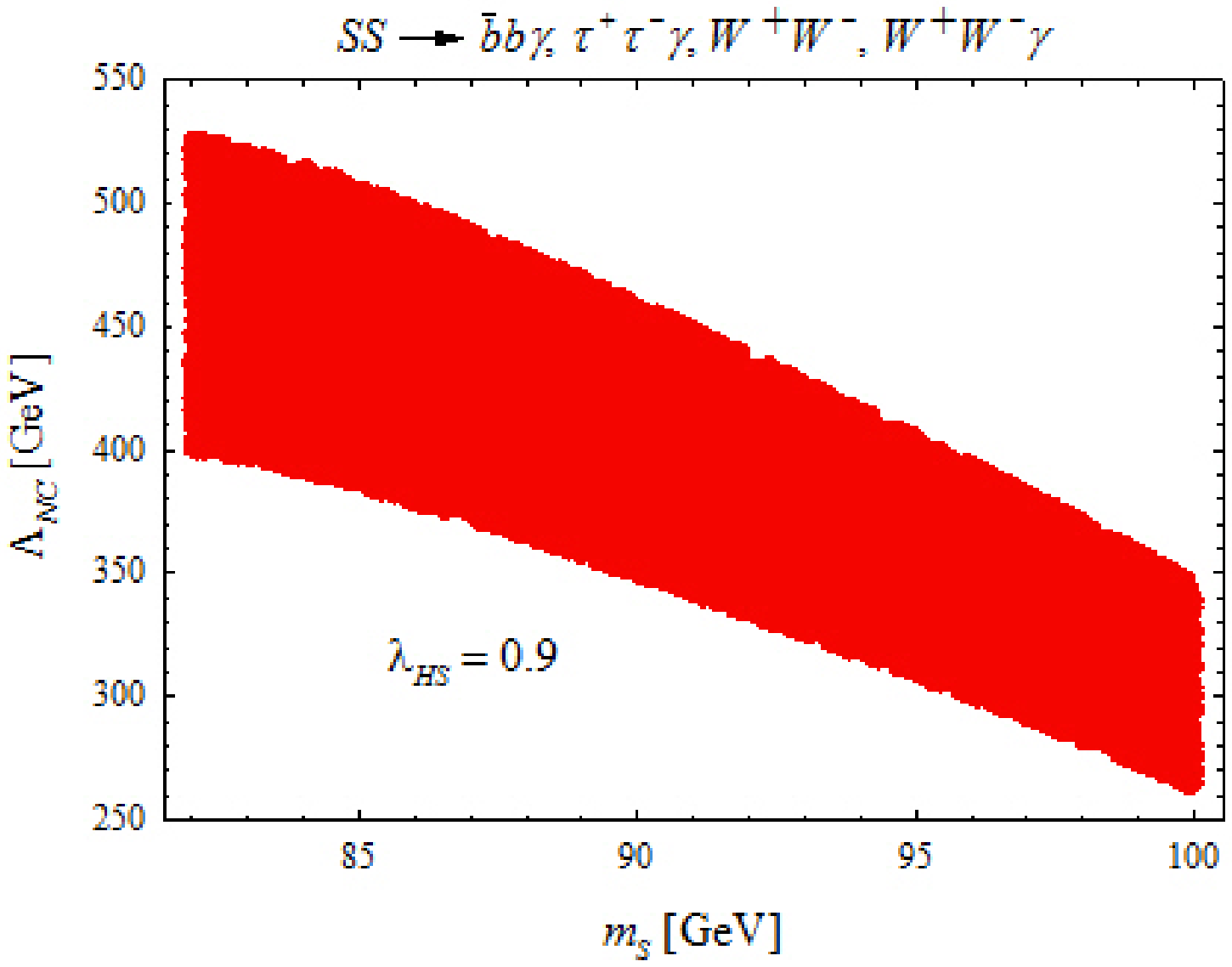}
\includegraphics[width=.30\textwidth]{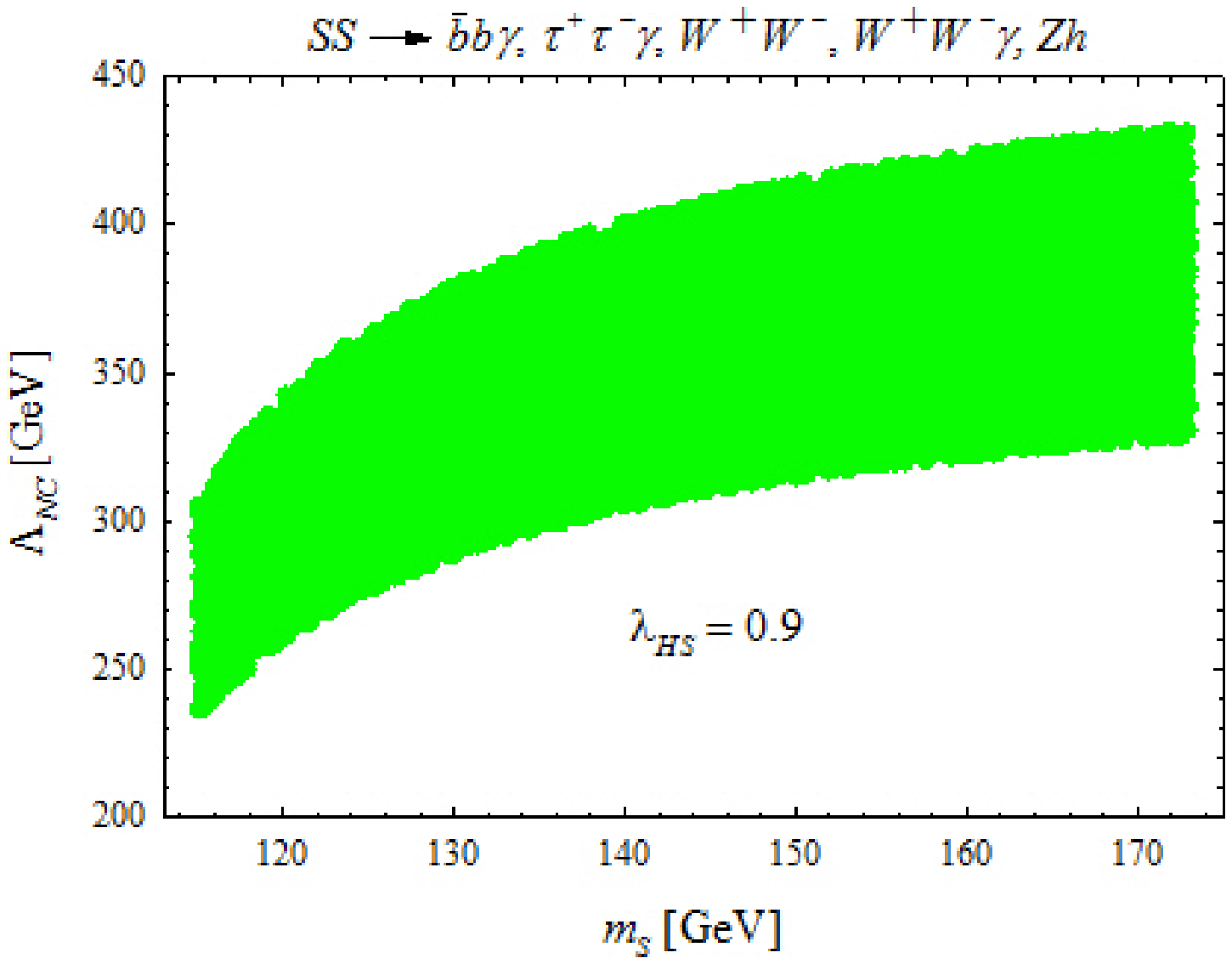}
 \includegraphics[width=.30\textwidth]{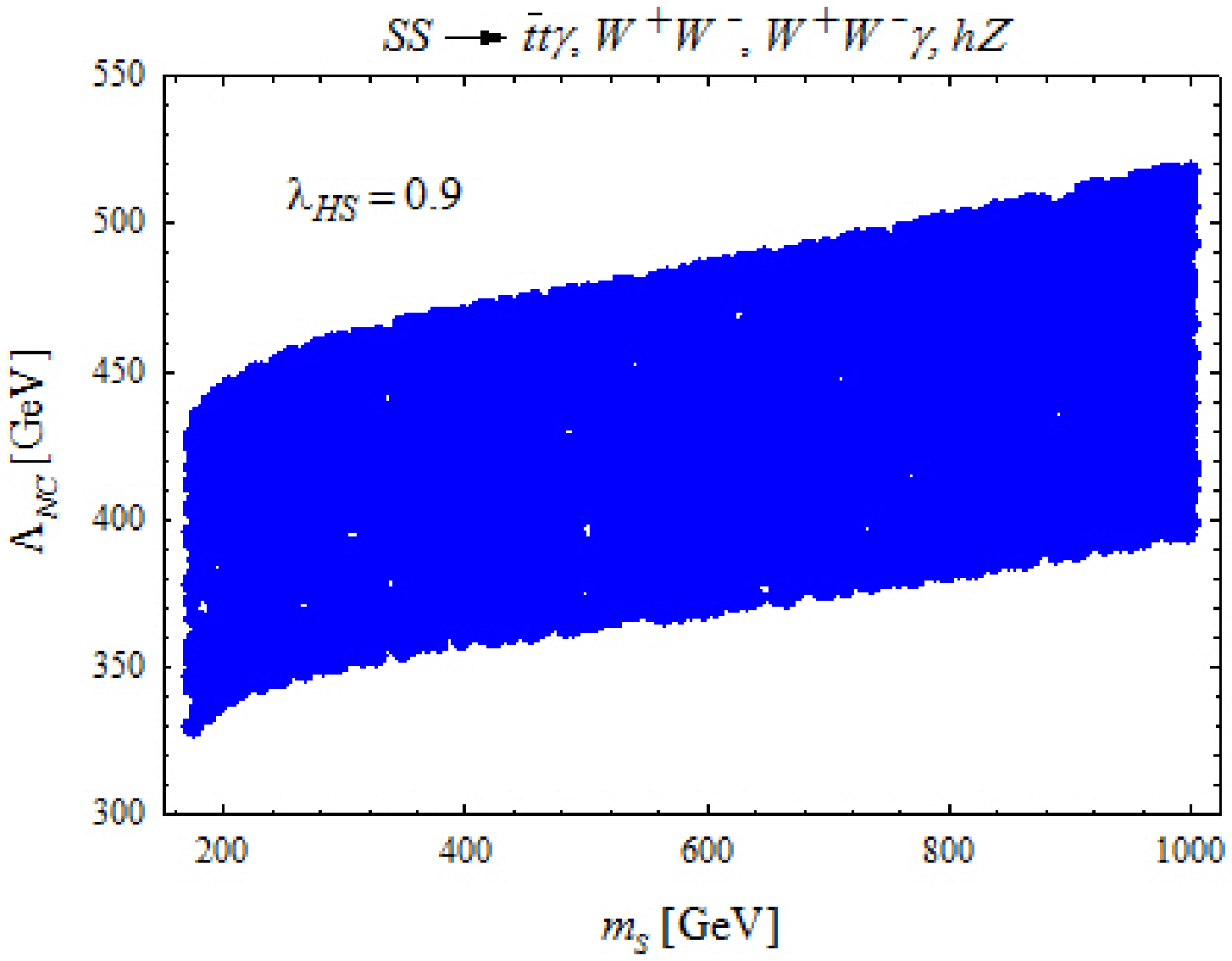}}
\centerline{ 
\includegraphics[width=.30\textwidth]{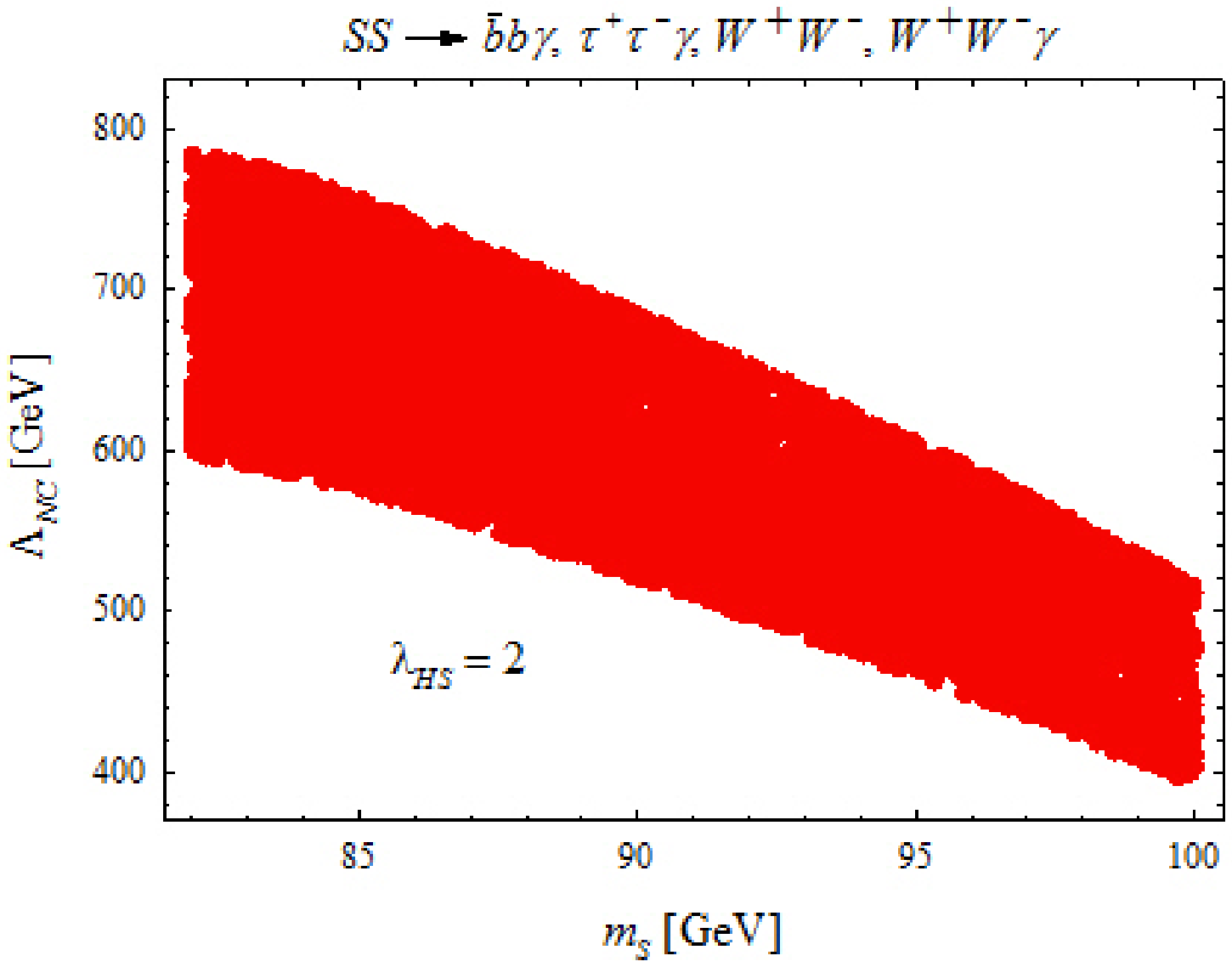}
 \includegraphics[width=.30\textwidth]{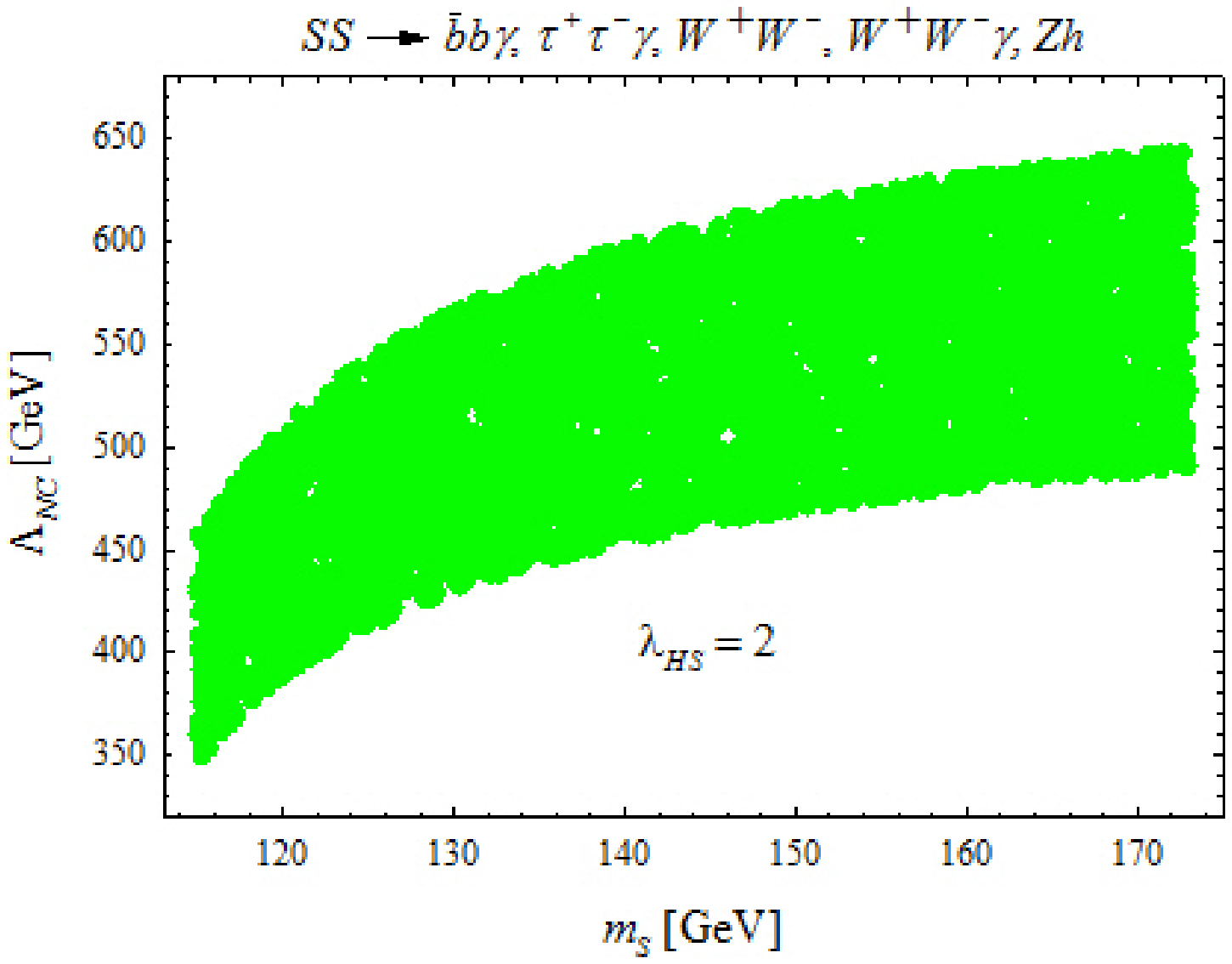}
\includegraphics[width=.30\textwidth]{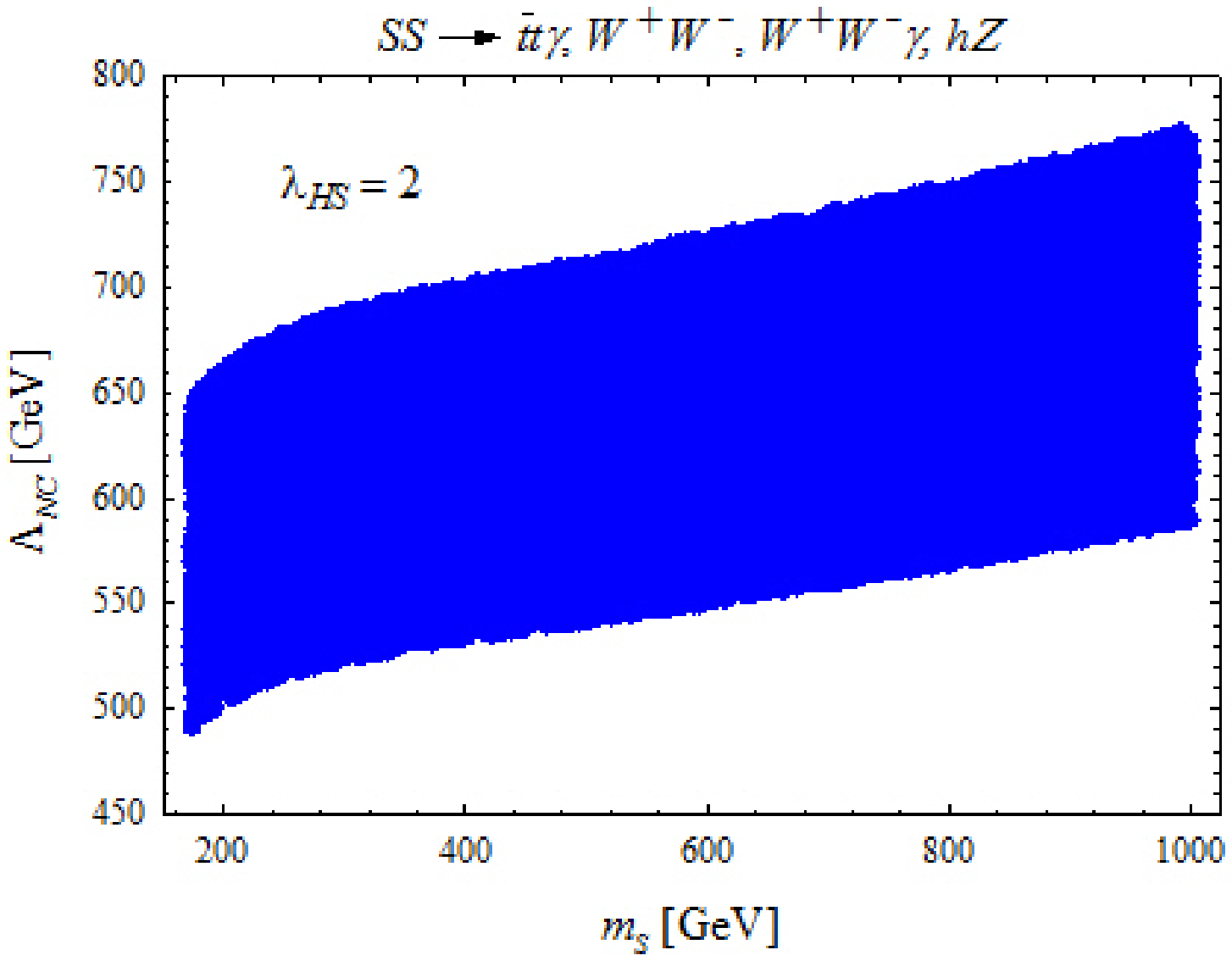}}
%\centerline{}
	\small\caption{The allowed region in the ($\Lambda_{NC},m_S$)-parameter space where two different values are adopted for $\lambda_{HS}$. The panels are labeled by their corresponding channels.}\label{fig.scatter-point2}
\end{figure}

We have performed our calculations with DM mass as it could generate the Higgs particle (around  $m_S \cong 62.5 $GeV), but we didn't obtain desired cross section.
%&&&&&&&&&&&&&&&&&&&&&&&&&&&&&&&&&&&&
As the observed relic density of DM is the most important aspect of DM phenomenology, we consider $(\lambda_{HS},m_S)$-plane in such a way that it could satisfy the experimental measurements \cite{Cline:2013gha}. The new $(\Lambda_{NC}, m_S)$-plane is also considered to investigate the NCST effect on the parameter space.
%In this way, we refer to \cite{} in which there are exact limits on scalar singlet DM from LHC probes. These limits come from the relic abundance, invisible Higgs decays and DM researches in direct and indirect experiments. These considerations put tight constraints on the mass and coupling ranges of DM where in the mass one, we have employed the related processes contributing in DM annihilation. 
The scatter points in the $(\lambda_{HS}, m_S)$-plane and $(\Lambda_{NC}, m_S)$-plane are depicted in Figs.~\ref{fig.scatter-point1} and \ref{fig.scatter-point2}, respectively. In Fig.~\ref{fig.scatter-point1}, the process is incremented for the first range of mass and is decremented for two others, for a fixed $\Lambda_{NC}$. $(\Lambda_{NC}, m_S)$-plane is depicted for two different values of $\lambda_{HS}$ (see Fig.~\ref{fig.scatter-point2}). The processes are the same for both $\lambda_{HS}$, but the difference goes back to the $\Lambda_{NC}$ amounts; the bigger amount of $\lambda_{HS}$ is, the larger amount of $\Lambda_{NC}$ it is.% and vice versa. 

%***********************************************
%***********************************************
 \begin{figure}[!htp]
	\centerline { 
			\includegraphics[width=.50\textwidth]{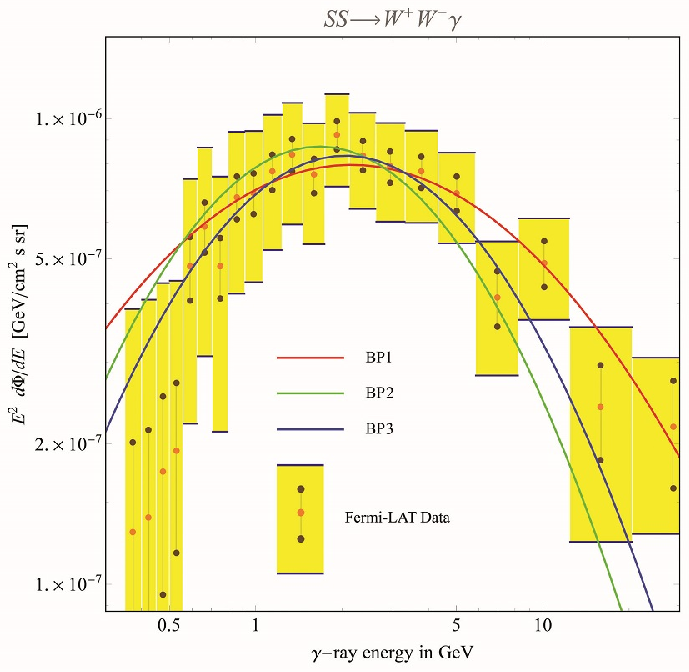}
                               \includegraphics[width=0.50\textwidth]{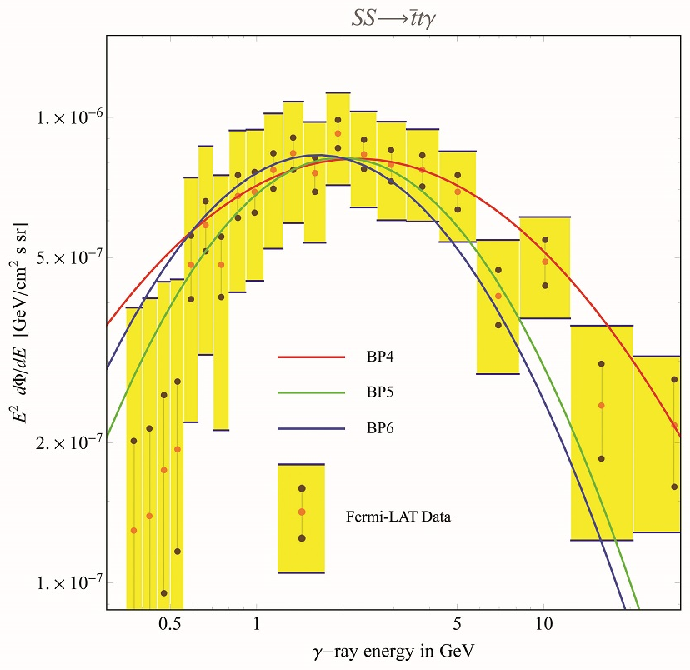}}
\centerline {\includegraphics[width=0.50\textwidth]{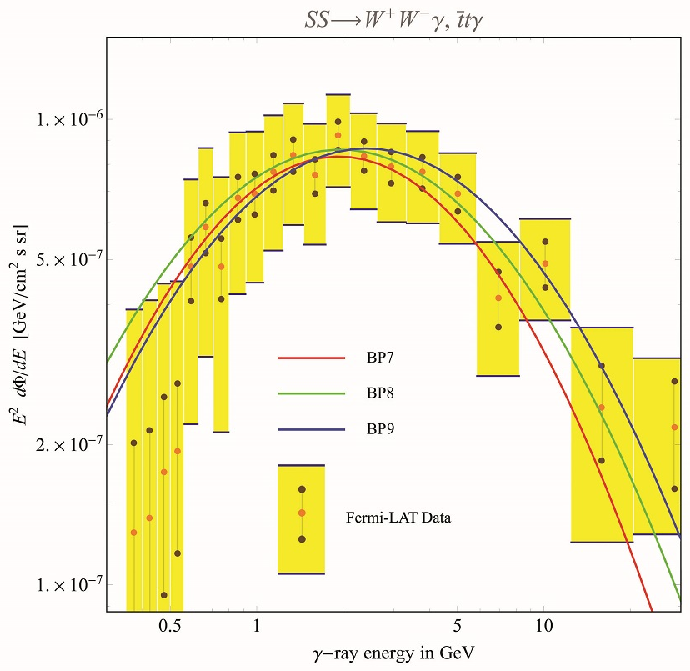}}
%\centerline{\includegraphics[width=0.50\textwidth]{Flux-SS--WWY, ttY.jpg} }
	\small\caption{Gamma-ray flux obtained from singlet scalar annihilation in the NCST in comparison with Fermi-LAT results. Allowed sets of parameters are presented in the form of nine BPs.}\label{fig.flux}
\end{figure}

%{\color{red}Now we are ready to construct our model parameter space following the existence experimental limit as ??. Thus we consider ranges of parameter space in the $\lambda_{HS}, m_S$-plane which can be embedded in the mentioned upper bound and lower one. Pursuing the above procedure, including relic density and mass constraint of DM, we probe our parameter space in three classes of benchmark points. Fig ? depicts our calculation probing a wide range of DM mass ending with massive DMs (Fig a-c).}
               
Explaining the gamma excess within the framework of singlet scalar DM is our main purpose in the presenting paper. Thus we analyze Fermi-LAT data via evaluating the gamma-ray flux in our model. As we mentioned earlier in Sec. \ref{sec:GR}, we adopt $\alpha\beta\gamma$ approach and NFW halo profile for DM using the appropriate Galactic parameters ($ |l|\!\leq\! 20^{\degree}\!$, $2^{\degree}\!\!\leq\! |b|\! \leq\!\! 20^{\degree}$). The photon flux for different benchmark points (BPs) is shown in Fig.~\ref{fig.flux}. Benchmark sets are tabulated in tables \ref{tab.1}$-$\ref{tab.3} where we have found the best fit model in the gamma-ray energy range of $1-3$ GeV. Considering the $SS\rightarrow W^+W^-\gamma$ channel at $\Lambda_{NC}=500$ GeV (upper panel in Fig.~\ref{fig.flux}), our best fit values include DM mass as $m_S=100, 300 , 500$ GeV and its coupling as $\lambda_{HS}=0.9, 1.6, 2$. % In this analysis, the best fit spectra are for fixed energy $E_b=100$ GeV accompanied with relevant $\alpha,\beta$ values  given in Table \ref{tab.1}. 
The upper panel of Fig.~\ref{fig.flux} shows the comparison of model flux with Fermi-LAT data using BP1, BP2 and BP3, where the best fit spectra are for fixed energy $E_b=100$ GeV accompanied with relevant $\alpha,\beta$ values  given in Table \ref{tab.1}. 

Our next analysis devotes to the DM annihilation into $\bar{t}t\gamma$ and the NC energy scale is chosen as $\Lambda_{NC}=500$ GeV. The best fit model is again categorized into three BPs (BP4, BP5 and BP6) including different DM masses and couplings where they are chosen as $m_S=200,500,1000$ GeV and $\lambda_{HS}=3$ depicted in the middle panel of Fig.~\ref{fig.flux}. %As can be seen in Table \ref{tab.2}, the best consistency of our model is at $E_b=300$ GeV where appropriate $\alpha,\beta$ are presented.
Appropriate $\alpha,\beta$ are presented in Table \ref{tab.2} where the best consistency of our model is at $E_b=300$ GeV.

Considering two aforementioned contributions in DM annihilation, now we investigate $SS\rightarrow W^+W^-\gamma, \bar{t}t\gamma$ channels to explore the model flux (lower panel of Fig.~\ref{fig.flux}). Comparing with Fermi-LAT data, we reach to the model best fit given in Table \ref{tab.3}. In BP7, BP8 and BP9 sets, $m_S$ ranges as 500, 700, 1000 GeV and $\lambda_{HS}$ reads as 0.9, 1.6, 2 respectively, for $\Lambda_{NC}=500$ GeV. Log-parabola spectrum indicates the parameters $\alpha,\beta$ with appropriate values at fixed energy $E_b=150$ GeV. 

In all above analyses, it is obvious from Figs.~\ref{fig.flux} that the best consistency occurs at the energy range of 1-3 GeV for gamma-ray emission in a centrally peaked form same as the reported map of the excess spectrum in the GC. It should be noted here that $SS\rightarrow\bar{b}b\gamma , \tau^+\tau^-\gamma$ channels are allowed but their contributions are small relative to the aforementioned ones (because of the small mass of $b$ quark and $\tau$ lepton against the $W$ boson and $t$ quark).
%%%%%%%%%%%%%%%%%%%%%%%%%%%%%%%

\begin{table}[!htb]
\caption{Bench mark points for  $SS\rightarrow W^+W^-\gamma $}
\label{tab.1}
\centering
\begin{tabular}{|c|c|c|c|c|c|c|}
\hline\hline
BP &  $\alpha$ & $\beta$ & $E_b$ & $m_s$ & $\lambda_{HS}$ & $\Lambda_{NC}$\\
     &        &      & GeV & GeV &     & GeV\\
\hline
$1$ & 2.6 & 0.48 & 100 & 100 & 0.9 & 500\\
\hline
$2$ & 4.1 & 0.9 & 100 & 300 & 1.6 & 500\\
\hline
$3$ & 3.87 & 0.85 & 100 & 500 & 2 & 500\\
\hline
\end{tabular}
\end{table}

\begin{table}[!htb]
\caption{Bench mark points for $SS\rightarrow \bar{t}t\gamma $ }
\label{tab.2}
\centering
\begin{tabular}{|c|c|c|c|c|c|c|}
\hline\hline
BP &  $\alpha$ & $\beta$ & $E_b$ & $m_s$ & $\lambda_{HS}$ & $\Lambda_{NC}$\\
     &        &      & GeV & GeV &     & GeV\\
\hline
$4$ & 2        & 0.24    & 300 & 200 & 3 & 500\\
\hline
$5$ & 4.115 & 0.74      & 300 & 500 & 3 & 500\\
\hline
$6$ & 4.69  & 0.85     & 300 & 1000 & 3 & 500\\
\hline
\end{tabular}
\end{table}

\begin{table}[!htb]
\caption{Bench mark points for  $SS\rightarrow W^+W^-\gamma, \bar{t}t\gamma$ }
\label{tab.3}
\centering
\begin{tabular}{|c|c|c|c|c|c|c|}
\hline\hline
BP &  $\alpha$ & $\beta$ & $E_b$ & $m_s$ & $\lambda_{HS}$ & $\Lambda_{NC}$\\
     &        &      & GeV & GeV &     & GeV\\
\hline
$7$ & 4.1 & 0.82 & 150 & 500 & 0.9 & 500\\
\hline
$8$ & 3.6 & 0.69 & 150 & 700 & 1.6 & 500\\
\hline
$9$ & 3.513 & 0.7 & 150 & 1000 & 2 & 500\\
\hline
\end{tabular}
\end{table}
%The best-fit model           is best-fit by                is also consistent 

%*****************************************************************************************

\section{Conclusion}\label{sec:con}
Fermi-LAT probes have opened a new window to pursue DM traces in the GC. Among a lot of theories suggesting the DM interpretation for this cosmic anomaly, we have evaluated singlet scalar particles as WIPM candidates for DM in the framework of the NCST. Extending the NCST beyond the SM, we could define new vertices which generate photon directly in the tree level of the SM channels. 
%through which DM particles annihilate away in SM channels. 
We have used this paradigm in order to investigate indirect signals of DM and have derived phenomenological constraints on singlet scalar DM features. First, we have calculated annihilation cross section and have depicted its behavior versus model independent parameters. In a complementary analysis, we have searched for allowed region constructed by specific model parameter space. Then we adopted a conservative approach to examine the photon flux reported by Fermi gamma-ray space telescope.

We considered all possible channels for DM annihilation in order to satisfy observational constraints. Dealing with DM-SM sector coupling at $\Lambda_{NC}=500$ GeV, for masses bellow $100$ GeV, $\lambda_{HS}$ varies from $0.9$ to $3$ with an increment dependence. In the region $120\leq m_S\leq 173$ GeV, $\lambda_{HS}$ decreases from maximum value $3$ to $1.2$ as a lower bound. For massive DM ($173\leq m_S\leq 1000$ GeV), we found a lower bound in DM coupling as $0.8$ and the upper one to be $2.08$ (see Fig.~\ref{fig.scatter-point1}).
In the $(\Lambda_{NC},m_S)$-plane, we found fits to the NC energy scale depending on DM mass. For low masses, $m_S\leq100$ GeV, $\Lambda_{NC}$ reaches up to $800$ GeV for interaction strength of $\lambda_{HS}=2$. Keeping up this $\lambda_{HS}$, the upper bound of NC scale changes to $\Lambda_{NC}=650$ GeV for $120\leq m_S\leq 173$ GeV and $\Lambda_{NC}=780$ GeV for $173\leq m_S\leq 1000$ GeV (see Fig.~\ref{fig.scatter-point2}). 

In the last analysis, we tested the viability of our model to explain the photon spectrum. For DM annihilation into $W^+W^-\gamma$, we have found three BPs in which the SM particles of mass $100,300,500$ GeV can generate sufficient gamma-ray flux. In the other pure channel, DM particles annihilate directly to photon via $SS\rightarrow\bar{t}t\gamma$ with DM masses as $200,500,1000$ GeV. % can feature larger interaction strength as $\lambda_{HS}=3$. 
Then a mixture of both channels ($SS\rightarrow W^+W^-\gamma,\bar{t}t\gamma$) was tested and the best fit values were for $m_S=500,700,1000$ GeV requiring coupling $\lambda_{HS}=0.9,1.6,2$ respectively.

Evaluating our results reveals that the NCST could be considered as a promising framework to increase DM annihilation cross section and validate indirect signals of DM detection. Although in this paper, we have investigated gamma-ray excess in the GC, the NCST can be considered as open research subjects in future DM phenomenological studies.  Our upcoming works will devote to the other phenomenological aspects of DM in the NCST (including other DM candidate fields such as fermiom, vector etc.). 

%******************************************************************************************************
\section*{ACKNOWLEDGMENTS}
We would like to thank Mohammad Mehdi Ettefaghi for reviewing the manuscript and we are particularly grateful to Manoj Kaplinghat, Can Kilic, Torbjorn Sjostrand, Marco Cirelli and Mohammadreza  Zakeri for useful discussions.
%\end{Acknowledgments}          

%******************************************************************************************************

\section*{APPENDIX}\label{sec:apx}
Here, we summarize DM annihilation cross sections presented in our calculations. We can write $\langle\sigma v\rangle_{i\to f}$ for the annihilation of DM of mass $m_S$ as
\begin{equation}\label{eq:ap1}
	\langle \sigma v \rangle_{i\to f} = \frac{1}{16m_{S}^{4}T K^{2}_{2}(\frac{m_{S}}{T})} \int_{4 m_{S}^{2}}^{\infty}  s \sqrt{s - 4m_{S}^{2}}  K_{1}(\frac{\sqrt{s}}{T})   (\sigma v)_{i\to f} ds,
\end{equation}
where $K_i (i=1,2)$ is the modified Bessel function and $T$ denotes the freeze-out temperature. As we have introduced new vertices in our hypothesis, using Eq. \ref{eq:ap1} may be severe and time consuming. Thus, we use the following expansion (for non-relativistic particles up to the second order) \cite{Ettefaghi:2009ai}
\begin{equation}\label{eq:exp}
	\langle \sigma v \rangle_{i\to f} \simeq  a^{(0)} + \frac{3}{2} a^{(1)} x^{-1} + \frac{15}{8} a^{(2)} x^{-2},
\end{equation}
where
\begin{equation}
	a^{(n)} =  \frac{d^{n}}{(d\epsilon)^{n}}\langle \sigma v \rangle_{i\to f} \vert_{\epsilon = 0},   \qquad \epsilon   =
	\frac{s - 4m_{S}^{2}}{4m_{S}^{2}},     \qquad         x^{-1} \equiv \frac{T}{m_{S}}.
\end{equation}
Following the above equations, we deduce that $a^{(0)} = a^{(1)} = 0$ and for $a^{(2)}$ we have 
\begin{equation}
	a^{(2)} = \frac{2 x}{K^{2}_{2}(x)}  \bigg( \frac{d}{d\epsilon} [\sqrt{\epsilon} (\epsilon + 1)   K_{1}(2 x  \sqrt{\epsilon + 1})]  (\sigma v)_{i\to f}  \bigg) \bigg\vert_{\epsilon = 0}.
\end{equation}
Substituting $a^{(0)},a^{(1)},a^{(2)}$ in Eq. \ref{eq:exp}, we obtain the following form for DM annihilation cross section
\begin{eqnarray}
	\langle\sigma v\rangle_{i\to f}=\frac{15}{4xK_2^2(x)}\Big(\frac{d}{d\epsilon}\Big[\sqrt{\epsilon}(\epsilon+1)K_1(x)(2x\sqrt{\epsilon+1})\Big](\sigma v)_{i\to f}\Big)|_{\epsilon=0}.
\end{eqnarray}
All the contributing cross sections presented in $(\sigma v)_{i\to f}$ follow as  
\begin{eqnarray}
	(\sigma v)_{i\to f}=(\sigma v)_{SS\rightarrow W^+W^-}+(\sigma v)_{SS\rightarrow hZ}+(\sigma v)_{SS\rightarrow f\overline{f}\gamma}+(\sigma v)_{SS\rightarrow W^+W^-\gamma},\nonumber\\
\end{eqnarray}
%%%%%%%%%%%%%%%%%%%%%%%%%%%%%%%%%%%%%%%%%
and they are expressed as bellow
%%%%%%%%%%%%%%%%%%%%%%%%%%%%%%%%%%%%%%%%%
\begin{eqnarray}
	(\sigma v)_{SS\rightarrow W^+W^-}=\frac{\lambda_{HS}^2m_W^4}{8\pi}\frac{\sqrt{s-4m_W^2}/(s\sqrt{s})}{(s-m_{h}^2)^2+m_{h}^2\Gamma_{h}^2}\Lambda_{NC}^{-4}(s+m_{h}^2)^2(\frac{s}{4m_W^2}-2),\nonumber\\
\end{eqnarray}
%%%%%%%%%%%%%%%%%%%%%%%%%%%%%%%%%%%%%%%%%
\begin{eqnarray}
	(\sigma v)_{SS\rightarrow hZ}=\frac{\lambda_{HS}^2m_Z^2}{8\pi}\frac{(s-m_h^2)^2/(s\sqrt{s})}{(s-m_{h}^2)^2+m_{h}^2\Gamma_{h}^2}\Lambda_{NC}^{-4}(\frac{(s-m_Z^2+m_h^2)^2}{4s}-m_h^2)^{3/2},\nonumber\\
\end{eqnarray}   
%%%%%%%%%%%%%%%%%%%%%%%%%%%%%%%%%%%%%%%%%
\begin{eqnarray}
	(\sigma v)_{SS\rightarrow f\overline{f}\gamma}&=&\int_{0}^{1}\int_{0}^{1}\int_{0}^{1}\int_{0}^{1}\int_{1}^{\frac{s-4m_f^2}{2\sqrt{s}}}\frac{|{\cal M}_{SS\rightarrow f\overline{f}\gamma}|^2}{F}\nonumber\\
	&\times&\frac{1}{(4\pi)^3}\sqrt{1-\frac{4m_f^2}{s-2\sqrt{s}E}}2EdEdx_2dx_3dx_4dx_5,
\end{eqnarray} 
%%%%%%%%%%%%%%%%%%%%%%%%%%%%%%%%%%%%%%%%%
\begin{eqnarray}
	(\sigma v)_{SS\rightarrow W^+W^-\gamma}&=&\int_{0}^{1}\int_{0}^{1}\int_{0}^{1}\int_{0}^{1}\int_{1}^{\frac{s-4m_W^2}{2\sqrt{s}}}\frac{|{\cal M}_{SS\rightarrow W^+W^-\gamma}|^2}{F}\nonumber\\
	&\times&\frac{1}{(4\pi)^3}\sqrt{1-\frac{4m_W^2}{s-2\sqrt{s}E}}2EdEdx_2dx_3dx_4dx_5,
\end{eqnarray} 
%%%%%%%%%%%%%%%%%%%%%%%%%%%%%%%%%%%%%%%%%
where $F=\sqrt{s(s-4m_S^2)}$ and the relevant amplitudes are given by
%%%%%%%%%%%%%%%%%%%%%%%%%%%%%%%%%%%%%%%%%
\begin{eqnarray}
	|{\cal M}_{SS\rightarrow f\overline{f}\gamma}|^2=\frac{4\lambda_{HS}^2e^2Q_f^2m_f^2}{(s-m_h^2)^2+m_h^2\Gamma_{h}^2}\frac{E^2(E_f^2-m_f^2)}{\Lambda_{NC}^4}(\sqrt{1-x_2^2})(\sqrt{1-x_4^2}),
\end{eqnarray}
%%%%%%%%%%%%%%%%%%%%%%%%%%%%%%%%%%%%%%%%%
\begin{eqnarray}
	|{\cal M}_{SS\rightarrow W^+W^-\gamma}|^2&=&\frac{8\lambda_{HS}^2e^2m_W^4}{(s-m_h^2)^2+m_h^2\Gamma_{h}^2}\frac{E^2}{\Lambda_{NC}^4}\Bigg\{\frac{4}{3}-\frac{2}{3m_W^4}\nonumber\\
	&\times&\bigg[(2E_W^2-m_W^2-E^2+EE_W)^2+2E(E_W-\frac{E}{2})(2E_W^2-m_W^2-\frac{E^2}{2})\nonumber\\
	&+&\frac{3}{2}(E_W^2-m_W^2)(2E_W^2+\frac{E^2}{2}-2EE_W)\sqrt{(1-x_2^2)(1-x_4^2)}\bigg]\nonumber\\
	&-&\frac{8}{3m_W^2}\bigg[(2EE_W-E^2-m_W^2)+\frac{3}{4}(E_W^2-m_W^2)\sqrt{(1-x_2^2)(1-x_4^2)}\bigg]\Bigg\}.\nonumber\\
\end{eqnarray}
%%%%%%%%%%%%%%%%%%%%%%%%%%%%%%%%%%%%%%%%%
Here, $e$ and $Q_f$ are the electric charge of electron and of fermion $f$ respectively. Energies $E_f$ and $E_W$ are also defined in terms of gamma-ray energy $E$ as
\begin{eqnarray}
	E_{f/W}=\frac{\sqrt{s}-E}{2}.
\end{eqnarray}
%%%%%%%%%%%%%%%%%%%%%%%%%%%%%%%%%%%%%%%%%

%\bibliography{SDM_NCSTBib}
%\begin{thebibliography}{99}

%\end{thebibliography}

\end{document}